\documentstyle[prc,aps,preprint,epsf,epsfig]{revtex}
\tightenlines
\newcommand{\be}{\begin{eqnarray}}
\newcommand{\ee}{\end{eqnarray}}
\begin{document}

\draft

\title{
 Equation of State, Radial Flow and Freeze-out\\
in High Energy Heavy Ion Collisions
}
\author{C.M. Hung and E. Shuryak  }
\address{
  Physics Department, State University of New York at Stony Brook,
 NY 11794-3800}
\date{\today}
\maketitle    

\begin{abstract} 
We have shown that recent experimental data on radial flow, both from
AGS and SPS energies, are 
in agreement with the Equation of State (EOS) including
the QCD phase transition.     
 New hydro-kinetic model (HKM) is developed, which incorporates
hydrodynamical treatment of expansion and proper kinetics of the 
freeze-out. We show
that the  freeze-out surfaces for different secondaries and different 
collisions are very different, and they are not at all isotherms
$T=const$
(as was assumed in most previous hydro works).
Comparison of HKM results 
with cascade-based event generator RQMD is also made in some
details: we found that both EOS and flow are in rather good agreement,
while the space-time picture is still somewhat different.

\end{abstract}  



 
\renewcommand{\thefootnote}{\arabic{footnote}}

\section{Introduction}

   One of the main physics goals of high energy nuclear collisions
includes a test of whether
 for $heavy$ enough ions  at the
   AGS/SPS energy range (10-200 GeV/A)  production of (locally) equilibrated
 hot/dense
hadronic matter really takes place. We do know that at the early
stages
of those collisions (few fm/c after the first impact) a very
large energy density of the order of several $GeV/fm^3$ is actually
reached.
How rapidly it is equilibrated and whether new phase of matter -
Quark-Gluon Plasma 
(QGP) - is indeed produced remains unclear.
One possible strategy
to answer those questions is relying on
 special rare processes happening at earlier
   stages,  the e/m probes \cite{Shu_78}), or  $J/\psi$
   suppression \cite{psi}. In both directions we have recent exciting
 experimental  findings \cite{CERES,NA50}.

  This work is however devoted to hadronic observables
related with production of the  usual  secondaries,
$\pi,N,K$ etc.  
 It is widely believed that their spectra
are not actually sensitive to questions mentioned above: and indeed, as
the produced multi-particle system  expands and cools, the
re-scattering  erases most  traces of the
   dense stage.
Nevertheless,  those which       are 
 $accumulated$ during the expansion remain,
 and thus provide valuable information
about the state of matter through its evolution.

  The central phenomenon of such kind discussed in this paper is
 a {\it collective  flow}. Its multiple studies
at Bevalac/SIS energies ($E/A\sim 1 GeV$) have shown a number of
interesting
effects. However,  it was  concluded that nuclear matter do not really
reach equilibration under such conditions. 

In contrast to that, in high energy
   pp (or even $e^+e^-$) collisions, the thermal description for
   particle spectra and composition
 works
surprisingly well \cite{Shu_74,Becattini}.
 At the same time, (except maybe at very high energies) 
there is $no$ observed collective radial  flow in these
cases  (see \cite{SZ} and the next section): 
this alone shows that the system is not truly macroscopic\footnote{  
 An explanation 
suggested in \cite{SZ} is that in pp/$e^+e^-$ collisions 
matter excitation is not strong enough to overcome the ``bag
pressure'', and small systems created have
 stabilised transverse size at some equilibrium value, and thus zero pressure.
Modern models explaining these data use
strings:  those
 are precisely such objects. Large systems created in nuclear
 collisions
must have positive pressure, and thus expand. }.

 In contrast to that, data for heavy ion collisions  show 
very strong flow. 
therefore suggesting that the excited system created do indeed behave
as
a truly macroscopic system. To test whether it is indeed so
is 
  the main $physics$ objective of this paper.
More specifically, we study whether available experimental data on
heavy ion collisions in AGS/SPS energy domain are consistent with (so
far
semi-qualitative) 
information
  about  the Equation of State
(below EOS) of hot/dense hadronic matter as  obtained
  from
current lattice QCD.

The central phenomenon studied in this work is
  the so called $radial$ (or axially symmetric)
flow, observed in central collisions. Current data are now rich enough to
allow systematic study of its collision energy and rapidity
dependence, as well as dependence on the nuclear size (A-dependence)
and the particular secondary particle involved. All those dependencies
are discussed below, and to a large extent reproduced
by our model.

  Another, more  $practical$ objective of this paper is to create
a {\it next generation}  model
for heavy ion collisions, to be called
Hydro-Kinetic Model (HKM). It incorporates
 three
basic elements of the macroscopic approach --
(i) thermodynamics of hadronic matter, (ii)  
hydrodynamics of its expansion, and (iii)
realistic hadronic kinetics at the
freeze-out. 
Most elements of the model have in fact been worked out in literature,
 some are new, but we think they are taken together for the first
 time.

 The hydro-based works available in literature\footnote{After a very
   long break, there was a workshop in Trento ECT, May 1997 devoted
entirely to this subject. Its proceedings (which will appear as a series of
papers in Heavy Ion Physics, 5 (1997).) should give rather complete
description of recent activities. } 
 aimed more at proper
parameterization of the initial conditions \cite{otherhydro},
which would then lead to $y,p_t$ spectra comparable with data. Among
recent papers let us mention \cite{frankfurt_3liq} which study the
first few fm/c and attempt to 
derive the initial conditions from 3-liquid model and  \cite{Schlei} which 
has studied some specific EOS-related observables. Probably the
closest in spirit to our work is recent paper \cite{MH}, in which 
the same  freeze-out conditions is used. 
Unfortunately, its  physical consequences are not studied in any
details,
and their method (referred to as ``global'' hydrodynamics) include
unnecessary averaging,
which significantly obscure them. 
To the extent we could trace them, 
 our findings actually qualitatively agree
with the results of \cite{MH}.
 In particular, we also found  that
the resonance gas EOS leads to too strong flow at SPS, while the
softer EOS including
the phase transition gives it about right. 

   Let us now comment  on the relations between our approach and
widely used  cascade ``event generators'' (Fritjof,Venus,RQMD,ARC
etc). Hydrodynamics
 and cascades
were often treated as alternatives, and many people trust
cascades much more, as those are ``based on known physics''.
 In reality, both 
 rather    should  be used as complementary tools.   

   The very fact that all event generators 
approximately work, in spite of huge differences between
them (their tables of cross sections, 
lists of resonances included, etc,  are different, some have strings
or even color ``ropes'') indicate that bulk
results are insensitive to those differences.
For example, particle
composition  appears to be rather well
equilibrated,
 explaining  insensitivity 
to details of the model in some observables.The simplest way to test
which  of those parameters are relevant is to
vary the input parameters: unfortunately, very little work was so far done
along this  line.
  Considering flow, one
should
obviously
have a look components of the  stress tensor, the
  pressure p and energy density $\epsilon$ .
Hadronic cascades (see \cite{sorge_equilibration} for RQMD) have 
 a very simple EOS,  $p/\epsilon\approx const$, typical to
thermal 
resonance gas (see below).

Obviously cascades have a lot to say about $later$ stages of the
collisions, at the so called freeze-out stage where interactions
stop, resonances decay etc. 
They also provide  more detailed
information (e.g. the degree of local chemical/thermal equilibration) 
which in principle\footnote{In practice, 
to our knowledge it was not even demonstrated
  that any of these  cascade codes satisfy the detailed balance
(say there are resonance decays and other
 2 to many hadronic processes without their inverse), and
that, even if given  time, they do lead to correct thermal equilibrium. 
} 
help us understand the validity limits of the macroscopic 
variables/approaches. We will discuss many of these issues below.

At the same time,  description based on hadronic cascade
of the $earlier$ stages of
the collisions  obviously has little  theoretical justification, and
fails in practice for sufficiently high energies (SPS).  
``Event generators'' therefore rely on specific models (color strings
and their
breaking, etc), introducing plenty of unknown
parameters or even concepts (e.g.``color ropes''). What is even worst,
these models have so far no connection to developments in
non-perturbative QCD, say to
 lattice studies of QCD
thermodynamics. They disregard such issues as chiral restoration
 and deconfinement, leading to disappearance the very objects they
 work with, hadrons and strings\footnote{ Therefore
 their phenomenological 
success is even used as an argument against reality of the QCD phase 
transition itself: needless to say,
we are strongly opposed to this point of view.}. 
Hydro description, on the other hand, is much simpler and 
operates directly with EOS, so in this framework
and can easily incorporate different scenarios (e.g., with or without
the QCD phase transition).

 Our last comment is practical: with experiments
proceeding from light ions to  heavy ones, and from the AGS/SPS to the RHIC/LHC
energies we have to deal with  many thousands of secondaries.
Direct simulation of all their re-scattering are neither practical not
necessary:
as soon as the system
are  much larger than the interaction range,
the system can be cut into parts (or
``cells'') which  evolve independently from each
other. 
  Furthermore, one may separate internal evolution (thermodynamics and
  kinetics) from cell's motion (hydrodynamics),
enormously simplifying the problem. 
 As multiplicities grow cascades become more and more
expensive, while the macroscopic approach  becomes only
more accurate: at some point going from one language to another
becomes inevitable.

The paper is structured as follows. In section 2 we
start with some phenomenological introduction into properties of the
radial flow, setting the problem to be discussed below. In section 3
we consider
thermodynamics of hadronic matter, using a rather standard model of
resonance gas plus QGP with bag-model EOS. The important step is
determination
of the particular paths the volume elements of matter make in the phase
diagram (e.g. temperature
T - baryonic chemical potential $\mu_b$) during expansion. Then we
determine 
effective EOS on these paths, to be used 
section 4 in hydro calculations.
  In this paper we would not discuss the non-equilibrium phenomena
  neither
at formation stage nor during the passage of the phase
transition. It is more important however to 
 address kinetic phenomena at the end of hydro expansion, the so called
 $freeze-out$ stage: this we do in considerable details in section 5.
 Here we separately discuss
chemical and thermal freeze-out and discuss how the final spectra of
secondaries are generated. Then we go to comparison of 
observables, and especially the radial flow, with experiment and
cascades
(RQMD), see section 6. Summary of the paper is contained  in 
 section 7.

\section{Flow: the phenomenological introduction }

   First of all, in order to put things into proper perspective and
introduce terminology, we remind that the collective 
flow can be observed as  follows.
(i) Axially
symmetric $radial$ and (ii) $longitudinal$ flow exist even for central
collisions. For non-zero impact parameter experiments have also 
seen  clear  signals for at least two non-zero harmonics in the angle
$\phi$, known as
(iii) $dipole$
and (iv) $elliptic$ flow.

 In this paper we study only the first of them, the radial
flow, so 
let us now comment on others. The longitudinal flow was studied a lot
in other hydro-based works\cite{otherhydro}: we decided not
to discuss  it here in details. It is a parameterization rather than 
a real prediction: the issue is obscured by an
 uncertainty in initial conditions. 

Asymmetric flow (iii) and (iv) is potentially
 very interesting,
especially the elliptic one \cite{Olli,Sorge_elliptic}.
The difference between the elliptic and
radial flow should  mostly appear due to earlier stages: it is obviously
an exciting subject for further work.
(At the moment we however feel that it is too early, 
one should be able to get more details from
 experiment first. For  recent
summary see \cite{exp_flow}). 

  The existence of radial flow in nuclear collisions was widely debated
 in literature for a decade. Phenomenological
 fits of the $p_t$
spectra of various secondaries by some (ad hoc) velocity profile (or even a
 single velocity value $v_t$) and the $same$  decoupling
temperature
 $T_f$ is possible, see  \cite{fit_hydro}.
Unfortunately, the
data allow for multiple fits, with wide margin for the trade-off between  
 $v_t$ and $T_f$. In particular, for heavy ions one can obtain
 equally good fits with
  ($T_f=140 MeV, v_t=0.4$) and  ($T_f=120 MeV, v_t=0.6$). 
However, as we will show below, the model used was 
oversimplified. Even the main assumption that  
one should  expect the same  $v_t$ and $T_f$ for all secondaries 
is in obvious contradition to
elementary kinetics of the freese-out.

\begin{figure}[h] 
\vskip 0.5in
\epsfxsize=3.8in
\centerline{\epsffile{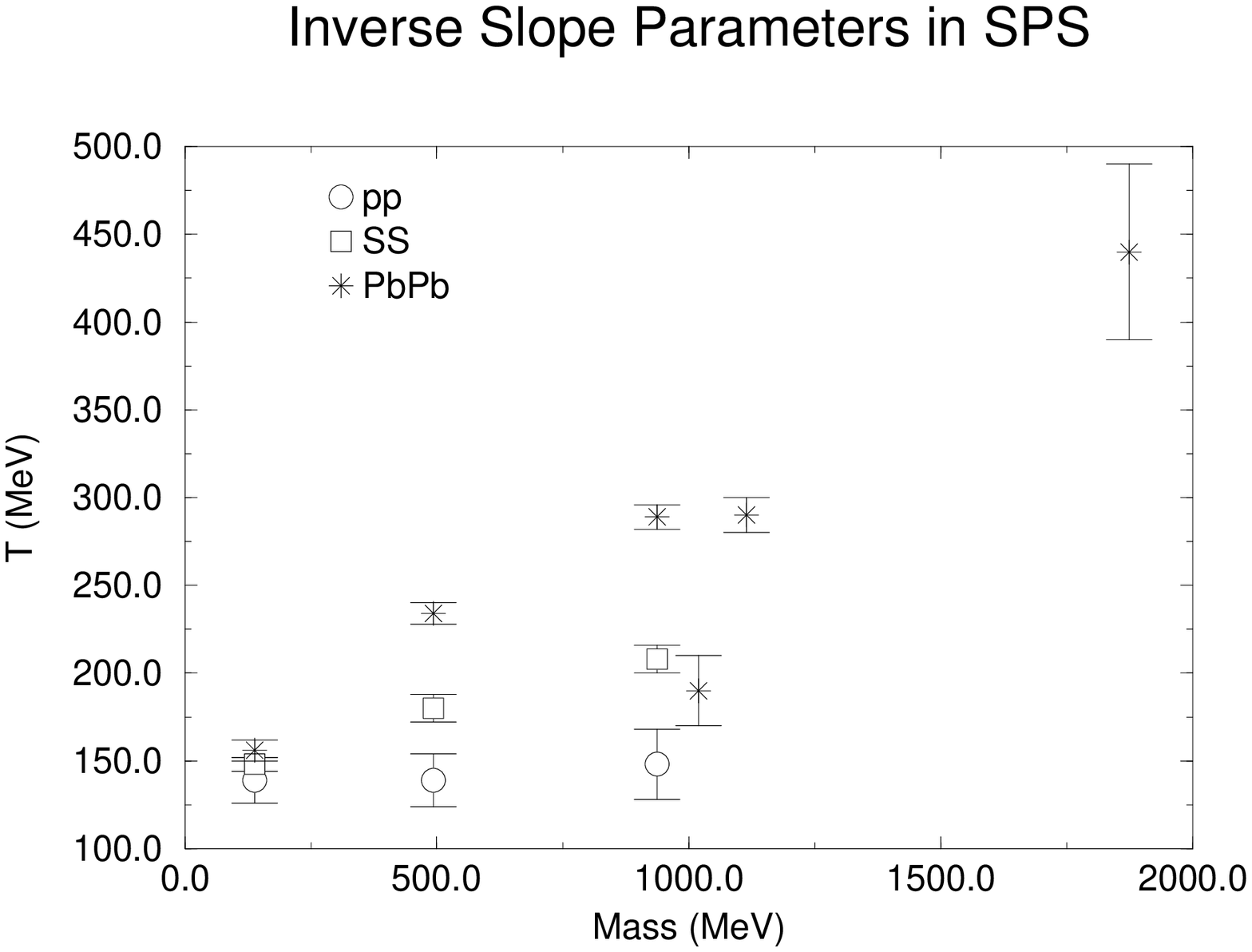}}
\caption{\label{mtslopes}
Experimentally measured slopes of $m_t$ distributions as a function of
particle mass (MeV), for $\pi,K,N$ (NA44) and $\phi,\Lambda, d$
(NA49), in acceptance of these experiments. Three
types of points correspond to pp, SS and PbPb collisions. }
\end{figure}

   More important is that rather rich experimental
    systematics is now emerging. In
Fig.\ref{mtslopes}
we show a collection of slopes from NA44 \cite{NA44_slopes}
and NA49\footnote{A disclaimer:  NA49 data
we use are preliminary  \cite{phi_NA49}, and is presented here
for qualitative comparison only. Also NA49 has  rapidity coverage wider
than NA44, and therefore their slopes should be somewhat
smaller. } experiments at SPS, for
 $\pi,K,N,\phi,\Lambda,d$. The definition is
\footnote{The tilde should remind that slopes are not temperatures,
 as they also include effect of the flow and resonance decays. }
\be E{dN\over d^3p} =C(y) exp(-{m_t \over \tilde T(y)}); \ \ & m_t^2=p_t^2+m^2 \ee

One major observation is a very strong {\it mass dependence}:
the
   slopes show consistent growth with the particle mass. 
It is clear how collective 
flow may explain it: for heavier secondaries, the thermal
motion is smaller and collective velocity $v_t$ starts to show
up. (Or, alternatively, collective motion  generates larger
momentum m$v_t$ for larger m.) Note however, that there are
no lines on this plot: as it will be clear from what follows, we
do not believe in any simple parameterizations, because 
participation in flow depends also on particle ability to interact
with others.

  An excellent example is the obvious exception from the general
  trend,
the  $\phi$ meson: 
 with about the same
mass as the nucleon, it has small slope.
 This is  because of {\it much smaller cross section
of its re-scattering},  leading to earlier decoupling,
so that $\phi$ does not participate in flow. (If, on the contrary,
the increased  slopes are  due to 
$initial$ state  scattering, as advocated e.g. in
\cite{Leonidov_etal},
one should instead get larger slope for $\phi$, since it is not stopped
by later ``friction force'' in matter.)

 Another excellent
test for existence of the flow 
is provided by deutrons.
 The shape of their  spectrum, its  slope $\tilde T_d$
and even yield are all very sensitive to
the  magnitude of flow.
For example, if flow is absent and
both protons and neutrons  be produced independently,
with a distribution $\sim exp(-p_t^2/2m_N T_N)$, their
coalescence into d would generate  distribution with the $same$
 $\tilde T_d=\tilde T_N$. The
observed  value is much larger.  The
 flow imply  a  specific
correlation between position and momentum, which helps to produce
larger $p_t$.
If this correlation is
artificially removed (see \cite{deutron_RQMD}, where
 in RQMD output the nucleon's positions or momenta were interchanged) 
the deutron spectra change shape and their yield  drops.

  The next point is storng A-dependence,  also quite evident from
 Fig.\ref{mtslopes}.
While the pp data show perfect thermal-looking spectra without a
slightest trace of the radial flow  \cite{SZ}, for SS collision the slopes
start growing with the mass of the secondary particle, and for PbPb
the effect is about twice larger. So,
the larger the nuclei, the stronger is the flow.

 This point is very important, because
such trend qualitatively contradicts to
what most of the  hydro models in literature would obtain.
The initial
longitudinal size is usually taken to be either
(i) the same for all, or (ii) it  scales as $A^{1/3}$.
With such assumptions
and  {\it A-independent freeze-out temperature}
 one  gets either
(i) a system which looks more and more one-dimensional, with
the radial flow   $decreasing$ with A, or (ii) a system with 
 a geometric scaling, with $A-independent$ flow.
Correct flow ($increasing$ with A)  naturally
follows from improved
 freeze-out conditions to be discussed below.

\begin{figure}[t]
\vskip 0.5in
\epsfxsize=3.8in
\centerline{\epsffile{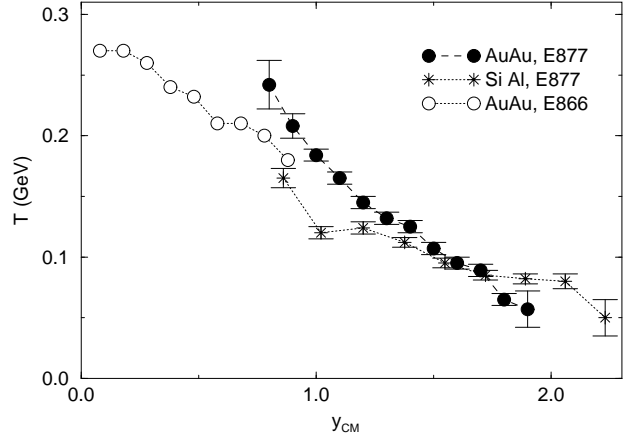}}
\caption{\label{AGSslopes}
Experimentally measured proton slopes of $m_t$ distributions at AGS
 as a function of rapidity y (counted from CM).}
\end{figure}

  One more aspect of the systematics
of the radial flow  is their $rapidity$ dependence. 
The nucleon slopes (taken from \cite{Sta_96})
from E877 and E866 experiments at   AGS  are compiled
  in Fig.\ref{AGSslopes}.
 One can clearly see from it that strong flow (and presumably its A-dependence)
 comes preferentially from
the central region, $y\sim 0$ in CM. 

  Compared to these strong trends, the observed dependence of the flow  on
the {\it collision energy} 
appears to be weak. Unlike for Bevalac/SIS energies, in which $v_t$
steadily grows, for AGS (10-15 GeV/A) and SPS (160-200 GeV/A)
 domain the radial flow velocity is (inside uncertainties)
about the same.  It must be a mere coincidence, since both
the meson/baryon ratio, the EOS and even the general picture of
 space-time development of the collisions are radically different.
Furthermore, as we have shown previously  \cite{HS_96},
 hydrodynamics predicts
 a rather non-monotonous dependence
of the lifetime of the excited matter as a function of the collision
energy,
with a sharp maximum between AGS and SPS energies.
This long time is related with a rather specific ``burning pancake''
regime:
and although no detailed calculation of radial flow was made, it is
hard to see how it can avoid having some kind of of discontinuity as well. 
 The
simplicity\footnote{
The EOS had no baryon number and full stopping was
assumed, which may not to be the case even for heaviest nuclei.}
 of the model used in \cite{HS_96} 
 somewhat limited its predictive power,
we believe the main conclusion about the peak of the lifetime should persist.
 One can look for
this effect experimentally by 
scanning to  lower energies at SPS, or by scanning various impact
parameters. (Although we have not studied non-central collisions in
this work, it is probably worth mentioning that such scan 
for $J/\psi$ suppression have shown discontinuous behavior at about
the same energy density.) 

  Another interesting manifestation of the ``softness'' of EOS is
  stabilization
of the radial flow at much higher RHIC energies. In this case hydro
predicts a ``burning log'' picture \cite{RG_96}, leading to mixed
phase surviving for 25-30 fm/c.  
As we will show shortly, this regime actually appears at SPS energies already.

\section{Thermodynamics of hadronic matter}

\subsection{Quark-gluon plasma  }

  Unlike  real experiments,  numerical ones performed on the
lattice are easier to do at higher T. As a result, 
current lattice data 
have significantly clarified the QCD thermodynamics of the quark-gluon plasma
phase. Above the phase transition region (see below) the
 thermodynamics was found to be
close to that of the ideal quark-gluon gas. Deviations are typically
10-15 percent downward shift in pressure and energy density
$p,\epsilon$  \cite{Laerman},
which are roughly reproduced by the lowest order ($O(g^2)$)
perturbative corrections\footnote{All higher orders
which can be perturbatively calculable have been now calculated,
  but those show divergent (or at least non-convergent) series, with
large and alternating sign terms.   }. Since for hydro only the
$p/\epsilon$
 ratio
 matters, this common factor can safely be ignored.
The non-perturbative corrections are more important: they are well
seen in lattice data for $T=(1-2) T_c$. Following tradition, we
parameterize it simply by addition of the bag-type term B
to the EOS of ideal quark gluon plasma
\be \epsilon = {\pi^2 T^4 \over 15}(16+{7\over 8} 6N_f)+{3N_f\over
   2}(T^2\mu_b^2 +{\mu^4\over 2 \pi^2}) +B \nonumber
\\
 p=  {\pi^2 T^4 \over 45}(16+{7\over 8} 6N_f)+{N_f\over
   2}(T^2\mu_b^2+{\mu^4\over 2 \pi^2}) -B
\ee   
The value of B is tuned to get $T_c=160 MeV$ for zero baryon density
(see below) resulting in
the value\footnote{ 
Note that it is about 6 times the original  constant of the MIT 
bag model, and also only about a 1/2-1/3 of what one would get if
all gluon condensate would be eliminated. }
 $B=320 Mev/fm^3$.

  Lattice data are also displaying a very spectacular phase transition in
  the  vicinity of $T_c$, in which $\epsilon$ grows by a
  large factor. 
 Although the exact dependence of order of the
  transition  on the theory parameters (such as quark
  masses, number of colors and flavors)
is still far from being completely clarified
(see \cite{Laerman,phases} for recent review), it already quite clear that
in practical sense 
the transition is  close to the 1-st order one with large
latent heat. Whether there is real jump, or just rapid rise
inside few MeV range of T can hardly be practically relevant:
 high accuracy of T cannot be reached for finite-size
systems we work with.

  The actually 
relevant variable  is not T but $\epsilon$: and
below we would refer to 
matter
 in wide
range of energy densities $\epsilon \sim .3-1.5 GeV/fm^3$
as a ``mixed phase'' domain. 
Its precise structure remains unknown: but
fortunately it should not matter for hydro,
provided the inhomogeneous domains (known also as ``bubbles''
of QGP) do not become too large. Fortunately, the hint we have
from lattice data is a predicted  tiny value (about 1
percent of $T_c^3$) for the surface tension. If it is true,
the  boundaries between the two phases cost little energy,
and so this phase should be very well
mixed indeed.

\subsection{Hadronic matter as a resonance gas }
   Ironically enough, properties of the hadronic matter at $T<T_c$
are theoretically understood much less than QGP.
 In many applications people usually simply
used the ideal pion gas as the simplest model: it leads than to a huge
latent heat in the transition.
  However this approach is clearly inadequate, and many more hadronic
  degrees of freedom are actually excited.

 We use instead the 
{\it resonance gas} approach, suggested very early by Landau and
Belenky
\cite{LB}. They have shown, using the lowest order virial expansion,
that  resonances\footnote{Those should be narrow enough: $\Gamma <<
  T$.}
 seen in scattering phases in fact
contribute to thermodynamical parameters exactly as stable particles.
It was later
used by Hagedorn in his statistical bootstrap studies of
 60's: his main point was  that the
$exponential$  mass spectrum leads to the upper possible
temperature
of the hadronic gas. 
  However, it was noticed by one of us long ago \cite{resonancegas} that
the observed resonance mass spectrum can better be fitted by power
of the mass than the exponent.
 It leads to rather simple EOS, for zero baryon number $p,\epsilon \sim T^6$
 \cite{resonancegas} or 
$p\approx 0.2 \epsilon$. Later much more detailed calculations with
actual scattering phases have
confirmed it.

  In this work we also include non-zero baryon density,
and so our thermodynamics have two variables, T and $\mu_b$ \footnote{The
chemical potential for strangeness $\mu_s$ is a $dependent$ variable,
with its value always fixed from the total strangeness S=0 condition.}. 
    Except at low energies (when we are close to nuclear matter), we
    know very little about the role of 
non-zero baryon density in the EOS. As it is
    well known, lattice calculations are so far impossible in this
    case, due to complex weight function for the non-zero chemical potential. 

  Simple generalization of the resonance gas to the non-zero chemical
  potential is of course natural, but it 
is known to have a problem at low T/high density.
Naive fermi-gas for nucleons clearly overestimates the
 pressure of nuclear matter.  The  QGP with reasonable bag constant 
 cannot compete with it, and therefore a phase transition line
has a pathological behavior at $\mu>$0.8 GeV (see
dotted line at fig\ref{fig_EOS}(a)).
Following many others
(e.g. \cite{PBM_etal}) we have solved this problem
by the excluded volume correction, which effectively reduces the
baryonic pressure at high $\mu$.  Specifically we adopted the excluded
volume model in \cite{excludedvolume}, which is thermodynamically
consistent,
and is characterized by the canonical
partition function

\begin{eqnarray*}
\label{eq-b-eos-11}
Z^{excl}(T,\{N_i\},V) = \sum_i{Z(T,N_i,V-V_0N_i)\theta(V-V_0N_i)}
\end{eqnarray*}

from which

\begin{eqnarray*}
\label{eq-b-eos-8}
P^{excl}(T,\{\mu_i\}) = \sum_i{P_i^{ideal}(T,\mu_i - V_0P^{excl}(T,\{\mu_i\}))}
= \sum_i{P_i^{ideal}(T,\tilde\mu_i)}
\cr
\end{eqnarray*}

$V_0$
is the excluded volume, which we assume to be the same for all
fermions, while $V_0 = 0$ for bosons.

\begin{eqnarray*}
\label{eq-b-eos-9}
n_i^{excl}(T,\{\mu_i\}) & = & \left( {\partial P^{excl}\over\partial\mu_i}
\right)_{T,\{\mu_i\}\backslash\mu_i}
= { n_i^{ideal}(T,\tilde\mu_i) \over 1 +
  V_0\sum_j{n_j^{ideal}(T,\tilde\mu_j) } }
\cr
s^{excl}(T,\{\mu_i\}) & = & \left( {\partial P^{excl}\over\partial T}
\right)_{\{\mu_i\}}
= { \sum_j{s_j^{ideal}(T,\tilde\mu_j)} \over 1 +
  V_0\sum_j{n_j^{ideal}(T,\tilde\mu_j) } }
\cr
\epsilon^{excl}(T,\{\mu_i\}) & = & Ts^{excl}(T,\{\mu_i\}) -
P^{excl}(T,\{\mu_i\}) + \sum_j{ \mu_jn_j^{excl}(T,\{\mu_i\}) }
\cr
& = & { \epsilon^{ideal}(T,\tilde\mu_i) \over 1 +
  V_0\sum_j{n_j^{ideal}(T,\tilde\mu_j) } }
\end{eqnarray*}

and the excluded volume radius $r_0=0.7 fm$.
 We do it just for completeness of the phase
diagram:
however we have to stress that all our
 results are completely independent on what is happening in this
 corner
of the $T,\mu_b$ phase diagram, since all the paths we discuss (see
below) are far from it.

  Apart from the excluded volume factor in p (which is also log Z), we
  use
standard thermodynamical formulae for the ideal gas of hadrons, stable
and resonances. We use all resonances till m=2 GeV.
All other variables are obtained from
the pressure $p(T,\mu)$ by standard thermodynamical relations.

\subsection{Adiabatic paths in the phase diagram and the resulting EOS }

  Although the $T-\mu$ plane is rather convenient for the determination of
 the  thermodynamical parameters in both phases, the  mixed
phase domain is hidden behind the transition line. As it is well
known, in the mixed phase new thermodynamical variable is the fraction
f of the volume occupied by the QGP phase. Besides, as we will show
shortly, the cooling trajectory in  the  $T-\mu$
plane is rather complicated.

 If expansion of matter\footnote{Note that we do not
  discuss
compression stage here: it is not slow and therefore entropy is in
fact produced here.} is slow enough, the {\it entropy is conserved}. We assume
it in what follows. 
 If so,  
the conjugates to ($T-\mu_b$) pair -- the
entropy s and the baryonic density $n_b$ -- provide more natural
description. If those variables are used, the cooling paths would be
just straight lines, going from initial point toward the origin.
As the entropy per baryon ratio stays constant, the paths can be
marked by this ratio. 

\begin{figure}[t]
\includegraphics[width=3.in,angle=-90]{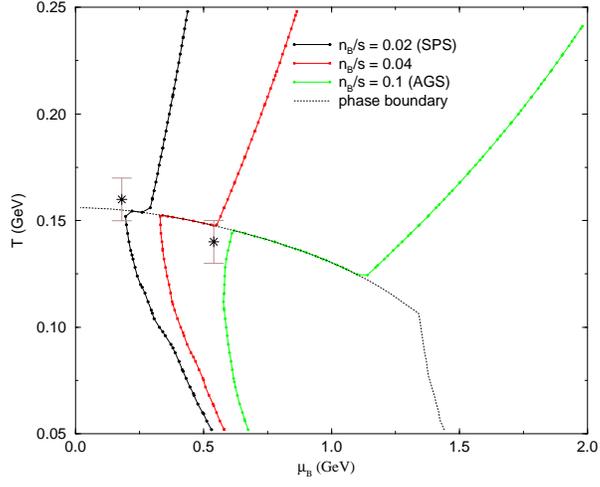}
\includegraphics[width=3.in,angle=-90]{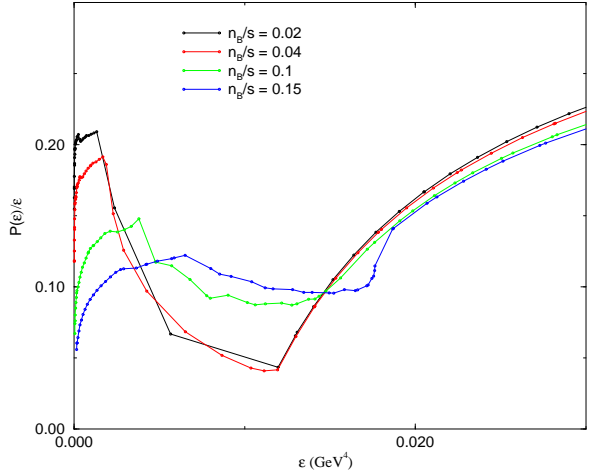}
\caption{\label{fig_EOS}
 (a) Paths in the $T-\mu$ plane for different
baryon admixture, for resonance gas plus the QGP; 
(b) the ratio of pressure to energy density $p/\epsilon$ versus
$\epsilon$,
for different baryon admixture. }
\end{figure}

   For the EOS described above
  (the {\it resonance gas}  for hadronic phase is supplemented by
a simple bag-type quark-gluon plasma)
we have calculated those paths in all variables. 
  	In Fig.\ref{fig_EOS}(a) we show how these
 paths look like on the original  phase
diagram. The lines are
marked by the $n_b/s$ ratio. Those for
$n_b/s=0.02,0.1$ correspond approximately
to SPS (160 GeV A) and AGS (11 GeV A) heavy ion
collisions,
respectively. Note that the trajectory has a non-trivial zigzag
shape\footnote{As far as we found, such shape first appeared in literature
  in \cite{Sub_86}.},
with re-heating in the mixed phase.
The endpoint of the QGP branch was named  \cite{HS_96} the ``softest
point'' , while the beginning of the hadronic one can be called the
``hottest
point''\footnote{Of course, in the ``Hagedorn sense'', as the hottest
  point of the hadronic phase.}.

  The next step is to define the effective EOS
in the form $p(\epsilon)$ (needed for hydro) {\it on these lines}:
that is shown in Fig.\ref{fig_EOS}(b)
.
Note that  the QCD resonance gas 
in fact  has  a very simple
EOS\footnote{The main difference between the curves with various
$n_b/s$ is at the
low energy density side: obviously adding baryons one contribute much
more to the energy density than to pressure. as we will show below,
it will have a significant impact on the mean radial flow.
}
 $p/ \epsilon\approx const$, while displaying strong dive toward the minimum
of  $p/\epsilon$ (the ``softest point'').
The contrast between ``softness" of matter at dense stages and relative 
``stiffness"  at the dilute ones is strongly enhanced for the SPS
case: it 
 is the main physical phenomenon we study below.
   
For comparison, one should also look at the (effective) EOS corresponding to
popular cascade event generators. For RQMD (with repulsive
potential between baryons) it was studied in recent
work
 \cite{Sorge_elliptic} for AGS energies. Rather simple EOS was found,
 about the same  for
 compression and expansion stages. For (transverse) pressure and
energy density it is approximately
 $p/\epsilon\approx 0.14 $. It is very close to what our resonance gas
gives for the corresponding $s/n_b$ ratio. (It would be nice to have
similar results for other cascades, and in wider energy range.)

  In summary: resonance gas (even with baryons) has a very simple EOS,
 $p/\epsilon\approx const(\epsilon)$. However lattice results
(modeled via bag-type model for QGP) indicate 
that EOS of hadronic
  matter is much {\it softer}, with  small  $p/\epsilon$ in the 
 interval of the energy densities near the end of the mixed
 phase. 

\section{Hydrodynamics}

  The equations of relativistic hydrodynamics are standard
\be \partial_\mu T_{\mu\nu}=0; \ \ & \partial_\mu n_b u_\mu=0
\ee
 In absence of any dissipative terms, they imply conservation of the
 entropy   $\partial_\mu s u_\mu=0$
and baryon number $N_b$.  The ratio of their local
  densities $n_b/s$ is not changing, and that is why 
in our discussion of the thermodynamics above we have parameterized by
it
the  paths
  on the phase diagram (e.g. $T,\mu_b$). Furthermore, we have shown
  that hydro-relevant form of EOS, namely $p(\epsilon)/\epsilon$,
  depends smoothly on this ratio.

  It was shown many times (see e.g. a recent review \cite{Sta_96})
 that\footnote{For clarity: this statements holds  for central collisions
of heavy enough ions, which
   have very small ``corona'' of punched through nucleons.
 For medium/light
    ions it is obviously more visible, and
asymmetric systems have many spectator nucleons.
Certainly those are not part of the hydrodynamic fireball. 
 },
for PbPb/AuAu collisions  at SPS/AGS energies the 
  rapidity spectra of  $\pi,K,N,d$
can be 
  described by 
 by some  $common$  collective motion, convoluted with thermal one
(and this is certainly different for $\pi,K,N,d$). It suggests that
 all matter elements have about the same   composition  
($n_b/s$). 

As it was explained in previous section, $n_b/s$ ratio is conserved
for each matter elements. However, if initial conditions have different $n_b/s$
in different places, it becomes space and time-dependent due to flow. 
 Phenomenological observations mention in the previous paragraph imply
that we may in fact significantly simplify the problem, assuming
``well mixed'' initial
conditions which have constant $n_b/s$ everywhere.
If so, equations for baryon flow and entropy flow become the same,
and  $n_b/s$ ratio is space-time independent.
   In practice, one can  determine  $n_b/s$
   from  the
baryon/meson ratio at the freeze-out stage. We use the values
$n_b/s=0.02, 0.085$ as representative
for SPS (160 GeV A) and AGS (11 GeV A) heavy ion
collisions,
respectively. These paths corresponding to them on the    $T-\mu$ plot
are also shown in fig\ref{fig_EOS}(a).

The initial geometry of the fireball was chosen to be Saxon-like with
natural Lorentz contraction in the longitudinal direction.
In this work we have
  not even attempted to discuss kinetics at the formation stage, and
  simply adopt a phenomenological approach, introducing
initial longitudinal size $z_0$ and velocity $v_z=v_0 tanh(z/z_0)$ 
as a phenomenological parameters. As a result, we do not
   have a predictive power as far as rapidity distribution
  is concerned: but we can just fit it (as it was many times done before, see
\cite{otherhydro}). We concetrate below at central region of rapidity, and
do not intend to describe well spectra in the target
or projectile fragmentation region. Although
we describe most of the secondaries,  the total energy
of hydro subsystem is
only a fraction of the total one.
  For 160A
GeV Pb+Pb the total initial energy of the hydrodynamical system is
about $0.4$ of the total center of mass collision energy
(which corresponds to initial central energy density of $4 {\rm GeV/fm^3}$),
while for
11.6A GeV Au+Au this ratio (the {\it inelasticity coefficient}) is about $0.7$ (which
corresponds to initial central energy density of $1 {\rm GeV/fm^3}$).

 The uncertain {\it initial conditions}
 are not important for transverse flow, because
it is accumulated over long time. 
  We will return to discussion of hydro and radial flow results later, after we
 study 
kinetics of the freeze-out in more details.

\begin{figure}[t]
\vskip 0.5in
\epsfxsize=6.in
\centerline{\epsffile{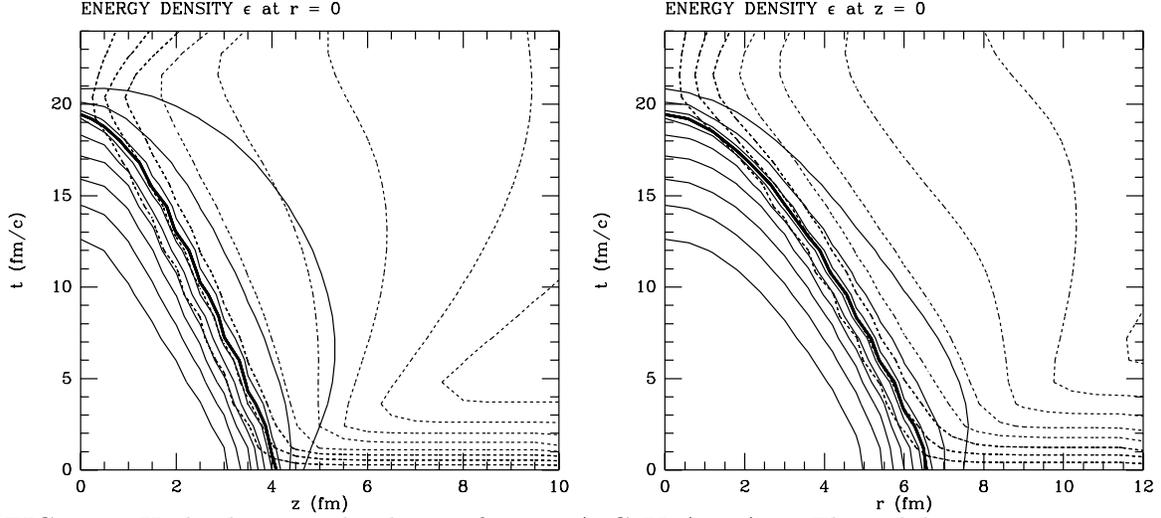}}
\caption{\label{fig-out210hydro}
Hydrodynamical solution for 11.6A GeV Au+Au. The solid contours are
energy density contours, with the bold contour being the
boundary between the mixed and hadronic phase ($\epsilon = 0.35 {\rm
  GeV/fm^3}$). The dotted contours are the longitudinal (left) and
radial (right) velocity contours, with values starting from the left
of 0.01, 0.05, 0.1, 0.2,...}
\end{figure}

\begin{figure}[t]
\vskip 0.5in
\epsfxsize=6.in
\centerline{\epsffile{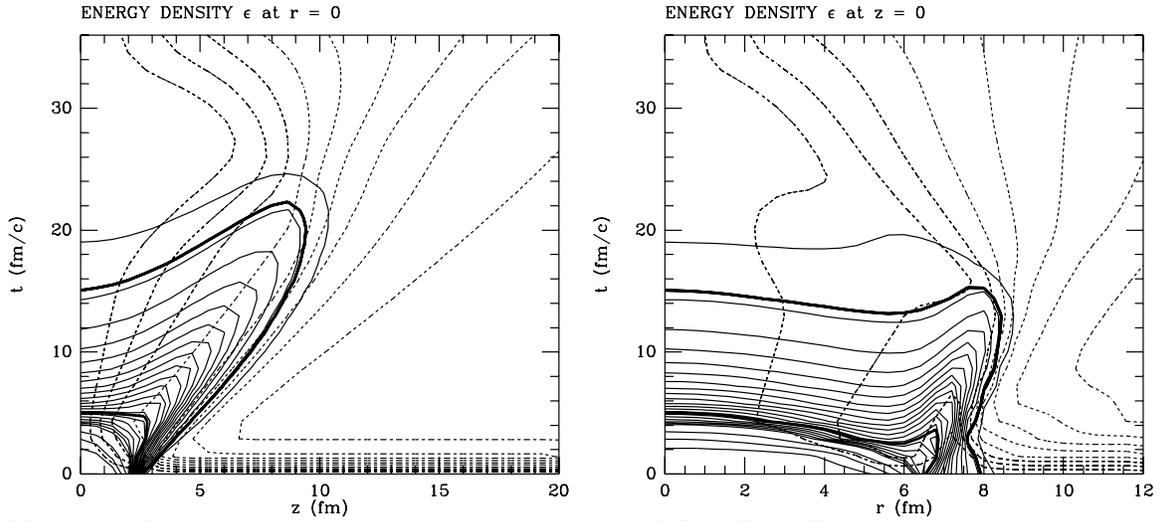}}
\caption{\label{fig-out213hydro}
Hydrodynamical solution for 160A GeV Pb+Pb. The solid contours are
energy density contours, with the bold contours being the
mixed-hadronic and quark-mixed boundaries, with energy densities
$\epsilon = 0.18, 1.4 {\rm  GeV/fm^3}$)
respectively . The dotted contours are the longitudinal (left) and
radial (right) velocity contours, with values starting from the left
of 0.01, 0.05, 0.1, 0.2,...}
\end{figure}

Typical solution for 11.6A GeV Au+Au is shown in
Fig.\ref{fig-out210hydro}
while for 160A GeV Pb+Pb it is shown in Fig.\ref{fig-out213hydro}.
Let us make few comments about them. First of all, they are
qualitatively different. While at the former (AGS) energy the
longitudinal and transverse expansion are not that different, at SPS
ones the $longitudinal$ flow has already distict ultrarelativistic
(Bjorken-like) features, with most isotherms being close to
hyperbolae, the
lines of constant proper time $\tau=\sqrt(t^2-z^2)$. What is less
obvious
(and follows from particular EOS including the QCD phase transition)
is also a dramatic difference in the $transverse$ flow at AGS and SPS
as well. The former can be described as ``burning in'', the lines of
constant
energy density moves invard with some small constant speed. At SPS
the mixed phase matter burnes into the low density hadron gas at a
``burning log'', which is  nearly time-independent and positioned
at transverse radius 6-8 fm. With time, as more matter flow from the
center,
there is even tendency to get by the end of the expansion a hole
at r=0, with less density there  than in the ``burning log'' region.
Such behaviour is a result of overshooting the ``softest point'' in
the intitial conditions, and it is even more dramatic at higher (RHIC/LHC)
energies, see \cite{RG_96}.

 (It is interesting to note, that our
 hydro solutions in many cases show late $implosion$, with subsequent
  secondary explosion from the center  r=0.
However, it is happening well after freeze-out, and thus
such hydro solution can only become physical if colliding
nuclei are much larger  than the heaviest existing ones.)

\section{Kinetics of the freeze-out}

Although  there is  rather substantial theoretical literature related with kinetics of
  freeze-out, and all major concept (and most of the details)
 to be used below has been
  developed before, 
  in most of previous  hydrodynamical models  the freeze-out
is formulated  in a very crude, oversimplified form.
 Most of them simply assume that
all reactions stop when the system reaches some
$universal$
``final
temperature'' $T_f\approx 140 MeV$. However this approximation
 is clearly inadequate since both (i) different processes have
different rates, say inelastic and elastic ones;
  (ii) different secondaries have different rates;
(iii) the expansion rates are very different for different
 colliding nuclei, and even for the same nuclei for different
matter elements.

  To learn more about freeze-out conditions and resolve
  the issue phenomenologically, one can 
  study various observables, such as HBT radii, deutron production,
  Coulomb effects \cite{Coulomb}, event-per-event fluctuations
  \cite{fluct}, the pion chemical potential  etc.
Except for the last one, we do not so far have HKM predictions for
them,
and live those for studies elsewhere.  

\subsection{Local freeze-out condition }
The central point we would like to make is as follows.
 Although 
 each individual matter  element follow roughly the same path
on the phase diagram, the relevant kinetics is $not$ the same
because they move along these paths with different
speed.
 In particular, 
the freeze-out happens at higher $T,\mu$ when time evolution
is faster (smaller initial system, or closer to the edge of the system)
relative to matter elements for which the evolution is slower.
Accounting for it turns out to be crucial for applications we
have in mind in this work.
 
Fortunately, after the global
collective motion of matter is already determined from hydro
calculation,  we know the expansion
rate
of any matter element at any time.
 With this information at hand, plus known 
 kinetics of various hadronic processes, we can
formulate realistic conditions under which 
subsequent $freeze-outs$ (decoupling of a particular reaction) take place.

   The principle idea of a freeze-out goes back
 to the famous 1951 Pomeranchuck paper
(which initiated  Landau to suggest a hydro-based
approach for the first time). The condition Pomeranchuck had in mind
is a relation between the mean free path and the system dimensions.  
The particular form  we use (as far as we know, mentioned first in
\cite{BGZ_78})  
 is based on  the similar condition, which is however $local$
(or differential) value of the ratio\footnote{
The non-local
condition in line with original Pomeranchuck idea is worked out 
in \cite{Grassi}, but we feel it is still way too complicated to use it
in practice in hydro context, because of the integrals toward future propagation
involved.
}
\be \label{eq_cond} \xi= \tau_{exp}/\tau_{coll} \ee of the
where $1/\tau_{coll}$ is a collision rate per particle considered per unit
 proper time.   
The invariant expression for the expansion time can be given in terms of the
4-velocity $u_\mu$ of the flow 
\be 1/\tau_{exp}=\partial_\mu u_\mu \ee  
while $1/\tau_{coll}$ is a collision rate per particle considered,
calculated in the cell proper time. 
Hydrodynamics is applicable
(the dissipative terms are small)   when $\xi>>1$,
 while if $\xi << 1$ the reactions in question can be ignored. The
 boundary
 at which $\xi \sim 1$ exist both at the
formation and expansion stages, forming some 3-surface around the 4-volume
in which hydro is applicable.
Furthermore,  in principle the situation is
more complicated, with
  $\xi$ large and small  for different variables.

  First of all,  
let us  distinguish two
 classes of reactions: (i) the inelastic reactions leading
to
creation or annihilation of a certain species of particles; 
 (ii) elastic
rescatterings  leading to simple momentum exchange.
It is well known that the former need higher collision energies than
the latter. (For example, in the gas of massless pions one can use
chiral perturbation theory to evaluate reaction rates, and pion production
depends on temperature as $1/\tau_{production}\sim
T^9$ while elastic re-scattering
is $1/\tau_{rescattering}\sim T^5$. Clearly, as the expansion cools the gas, their
decoupling happens at different points.) 
 Separating those two
classes,
one usually defines
  $chemical$ and $thermal$ freeze-out, for these two
classes
of reactions.
The second important point:
  both  freeze-outs should be determined  for $each$ species separately.

 For   chemical freeze-out
this distinction however is not very important in practice, since
  in fact  all reactions changing the particle composition can bee seen
to be rather  ineffective during the
  hadronic phase, for all  AGS/SPS
  collisions\footnote{For example,
 for strangeness production reactions
   this statement was well documented  
    long ago, see \cite{RKM}.}.
In QGP (most) hadrons do not exist at all, and thus
the natural place for ``hadronization'' is what we call the mixed phase.
How it happens remains unknown: but there are rather convincing
arguments that it happens rapidly enough.
Those are based on
  quite extensive work on thermal description of many hadronic species
  \cite{strangeness,PBM_etal}, mainly in connection
with the so called ``strangeness enhancement'' phenomenon.
It was found that (within the existing experimental uncertainties, not
always small) one can describe most of particle ratios in a thermal model.
(Moreover, for heavy ions one can do it even without any ``strangeness
suppression''
factors.) The resulting values for $T,\mu_b$ are shown  in
Fig.\ref{fig_EOS} as two crosses, for the AGS and SPS energies respectively.
Both are close to the ``hottest points'' of the corresponding paths:
this 
is consistent with the idea that      
chemical equilibration cannot indeed be kept in the
hadronic gas phase. 

 We have built in this idea into HKM:
  any ``hadronic chemistry''
in the hadronic phase is ignored. It is  
 assumed that  it ends together with hadronization, and
when the path departs from
the phase transition line $T_c,\mu_c$ no more changes in particle
composition (apart of resonance decays) are included.
 Therefore our 
particle composition is exactly the same as  in the thermal model 
  \cite{PBM_etal} (which has thermodynamics of exactly the same resonance gas
 with excluded volume). We therefore do not duplicate
the tables for particle ratios here, referring the interested reader
to this work.

\subsection{ Between chemical and thermal freeze-out}

 We do not provide extensive introduction  for this section: for
a good overview
and  references see
   \cite{Bebie_etal}.Switching off all reactions changing the particle composition, we have made
 $any$ particle number $N_i$ to be a conserved quantity.
The point is simply that at this stage of the evolution one has to introduce
 chemical potentials for all particle species\footnote{For
  clarity: those potentials are conjugated to  
total number of particles, so say for pions they enter distributions of
$\pi+,\pi^-,\pi^0$  with the same sign.}, $\mu_i$. Their
values  are then determined by those pre-determined 
values of $N_i$ in the usual way.
 This is in contrast to
 chemical equilibrium, in which most of them are zero, and only  chemical potentials
conjugated to conserved quantities (baryon charge and strangeness)
were needed.

 It is instructive to see how, as
   one starts with chemically equilibrated hadron gas
with  $\mu_i=0$, the non-zero values appear as the system cools  further\footnote{
To our knowledge, it was first pointed
out in the context of the pion gas by G.Baym. Further
discussion of this idea and of kinetics of the 
 pion gas can be found in
\cite{Gavin}, mostly in relation with the question of
the possible evolution of the
 non-zero  chemical potential for the pions.
For discussion of the $opposite$
  scenario, suggesting overpopulation and large positive chemical
  potential for pions already at this
  point, see \cite{pioverpopulation}}.
Thermodynamical
relation
written
in
the
form
\be
(\epsilon+p)/nT - s/n = \mu/T.
\ee
is
especially useful. For slow (adiabatic) expansion the 
s/n ratio is not changing, while in the first term in l.h.s. the  
chemical potential $nearly$ cancels (it does
provided Boltzmann approximation is used). So, one
can read T-dependence of  $\mu$ directly from the r.h.s.  
The  notorious exceptional case worth mentioning is that for 
massless
particles, for which the whole l.h.s is just a constant. Therefore
 $\mu/T=const$, and so if $\mu=0$ at the beginning
it remains so for any T\footnote{This is what happens in the case of
  background radiation in expanding Universe: photons do not collide
  after the Big Bang, but they still have the Plank spectrum, with
  $\mu=0$.
}.   

Accounting for the non-zero pion mass and Bose statistics one finds
$\mu_\pi(T)$, see  \cite{Bebie_etal}. For example, 
if one assumes that $\mu_\pi(Tc=160 MeV)=0$, one finds that by a thermal
freeze-out (which happens for PbPb collisions at CERN
at T=110-120 MeV) the pion chemical potential 
$\mu_\pi=60-80 MeV$.  

\begin{figure}[t]
\includegraphics[width=3.8in,angle=-90]{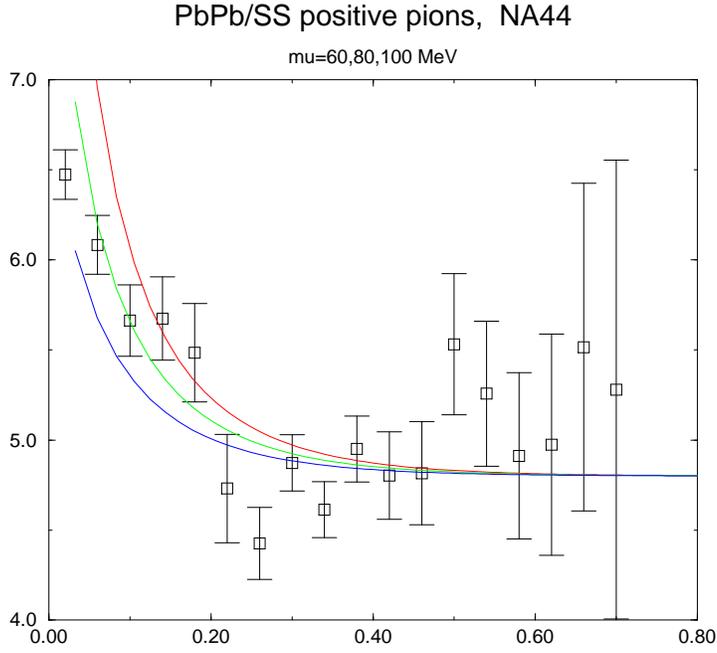}
\caption{\label{pionmu}
The ratio of $\pi^+$ $p_t$ spectra for   PbPb to SS
collisions. Points are experimental data from NA44 experiment,
three curves correspond to
 pion chemical potential $\mu_\pi=$ 60,80 and 100 MeV (from bottom up). }
\end{figure}

  In order to see whether such effect really occurs in experiment, we
  have plotted
in Fig.\ref{pionmu} the ratio of $p_t$ spectra for PbPb collisions (in
  which we expect  thermal
freeze-out at T=100-120 MeV, and thus formation of significant 
 pion chemical potential) to 
our reference point, central SS collisions (for which the effect should
 be much smaller). The data sets are both for $positive$ pions
\footnote{The $\pi^-/\pi^+$ ratio show larger enhancement, which is
known to be  due to Coulomb effects,
see e.g. \cite{Coulomb}. } from NA44 experiment \cite{NA44_slopes},  in the same experimental settings
(and thus systematic errors should somewhat cancel). One finds,
that there is significant enhancement of this ratio at small $p_t$,
which agrees  with 
formation of the non-zero  pion chemical potential. Moreover, as one
can see, the magnitude of the effect is in approximate agreement with our 
estimates. 

(Additional comments:
PbPb and SS collisions have somewhat different stopping of baryons.
 For $positive$ pions extra stopped charge for  PbPb  
would decrease low $p_t$ pion production due to the Coulomb field,
 contrary to observations. Another effect 
contributes in the opposite direction  is feeding to low  $p_t$ pions
 from extra $\Delta$ decays coming from extra baryons
 in PbPb as compared to SS. Magnitude of those effects is comparable,
and thus they may cancel out to some extent.)

 The
 secondaries other then pions can be to a good accuracy 
treated as Boltzmann non-relativistic gas, and so one can easily derive the
following
relation between the  chemical potentials at chemical and thermal
freeze-out
\be \mu_{th}=\mu_{ch}{T_{th}\over T_{ch}}+m(1-{T_{th}\over T_{ch}})      \ee
(In particular, for very large systems $T_{th}\rightarrow 0$,
 the chemical potential
 $  \mu_{th}\rightarrow m$ as it should, and one can then proceed to
 normal non-relativistic notations.) When implemented in HKM, this
 relation ensure that particle ratios are $independent$ on any details
 of thermal decoupling we discuss below.

\subsection{ Thermal freeze-out for different species}

  Now we are in the position to discuss particular reactions
in the resonance gas. Rather extensive studies have been made in the
past,
see \cite{PPVW_rates}. Let us start with qualitative comments first.

   Out of many reactions which include pions  the major processes are
the low energy  elastic
 $\pi\pi,\pi K$ and $\pi N$ scattering. Those  have especially large  cross
section
due to existence of the low energy  resonances, $\rho,K^*,\Delta$ respectively.

  The estimates of  $\pi\pi$ collision rate using the chiral
  Lagrangian
was made by one of us \cite{Shu_pions} and, in more details, 
in ref.\cite{GGL_89}. The result 
\be 1/\tau_{\pi\pi}=T^5/(12 F_\pi^4) \ee
display very strong T-dependence. This feature remains true when one includes
the resonances \cite{Bebie_etal}: basically in the interval
we deal with (T=120-150 MeV) the pion-pion scattering rate increases by
the factor 2\footnote{Furthermore,
 the inclusion of resonances changes
the dependence on the pion momentum p: in contrast to chiral result
 the rate becomes basically flat for $p<700 MeV$ we need, and decrease
 for larger p (now, in contrast to 
the lowest order chiral result which predicts the unphysical  
rise with p).} These rates are  increased further by the inclusion
of the
non-zero value of pion chemical potential 
discussed in the preceding subsection.

  Strong T-dependence leads to the following qualitative feature of
  freeze-out:
relatively modest changes in the freeze-out temperature correspond to
quite
 significant changes in duration of the collision-dominated (hydro)
expansion. As we will see below, this will translates to significantly
stronger flow.



  The $\pi N$ cross section is very large,
reaching about 200 mb at the $\Delta$ resonance peak.
Naive radius of the interaction  $R=\sqrt(\sigma/\pi)\approx 2.6 fm$
is so large that one may question simple cascades and think about
collective effects (``pi-sobars'').
Absolute scattering 
rates depends on the density of nucleons at
  the decoupling stage. At AGS 
 the (isospin averaged) rate is of the order of
 $1/\tau_{\pi N}\approx 100 MeV$, which is 
larger than   $1/\tau_{\pi\pi}$. Since nucleon to pion ratio is about
one, the rates are very close also.
At SPS energies the situation is quite different: the 
 nucleon/pion ratio is about 1/5. It  makes
the $\pi N$ scattering  less important for pions, but  nucleons have
very large collision rate and thus should freeze-out very late.

 Kaon and other strange secondaries have smaller collision rates. We have
 already mention a special case of
 $\phi$
 with the scattering and absorption cross sections 
in few mb range. Clearly one can completely
ignore their re-scattering in hadronic
phase: 
we assume therefore that their thermal freeze-out (as well as chemical
one) 
coincides with the end of the mixed phase.

  Let us now provide more quantitative information about
the rates we use (see also \cite{PPVW_rates}).
The general formula for the averaged collision rate of particle $a$
resulting from binary collision with particle $b$ is given by

\begin{eqnarray}
\label{eq-app-collision-6}
\Gamma^{a}_{ab}(T) =
{1\over{\int{d^3{\bf p}_a \over (2\pi)^3} {1\over e^{E_a/T} \pm 1}}}
\int{d^3{\bf p}_a \over (2\pi)^3}\int{d^3{\bf p}_b \over (2\pi)^3}
{1\over e^{E_a/T} \pm 1}{g_b\over e^{E_b/T} \pm 1}
\cr
\sigma_{ab}\left[(p_a + p_b)^2\right]
\left| {{\bf p}_a\over E_a} - {{\bf p}_b\over E_b} \right|
\end{eqnarray}

where $E_a$ is the energy of $a$ (minus any chemical potential for
$E_a$ within the thermal exponent), and
similarly for $b$.  $g_b$ is the multiplicity of $b$ and the sign in
the denominator of the thermal weights are chosen based on whether
$a$, $b$ is a fermion or boson.
For example, for
the $\pi N$ rate we take the $\pi^+p$ total
cross-section from Particle Data Group\cite{pdgpiN} and notice that by
isospin arguments, the averaged $\pi N$ cross-section is

\begin{eqnarray}
\label{eq-app-collision-7}
\sigma^\pi_{\pi N} \simeq {2\over 3}\sigma_{\pi^+p}
\end{eqnarray}

also noting that $g_N = 2$, we get for the $\pi N$ pion collision rate

\begin{eqnarray}
\label{eq-app-collision-8}
\Gamma^{\pi}_{\pi N}(T) =
{1\over{\int{d^3{\bf p}_\pi \over (2\pi)^3} {1\over e^{E_\pi/T} -1}}}
\int{d^3{\bf p}_\pi \over (2\pi)^3}\int{d^3{\bf p}_N \over (2\pi)^3}
{1\over e^{E_\pi/T} -1}{4/3\over e^{(E_N - \mu_b)/T} +1}
\cr
\sigma_{\pi^+p}\left[(p_\pi + p_N)^2\right]
\left| {{\bf p}_\pi\over E_\pi} - {{\bf p}_N\over E_N} \right|
\end{eqnarray}

\begin{figure}[t]
\begin{center}
\includegraphics[width=3.8in]{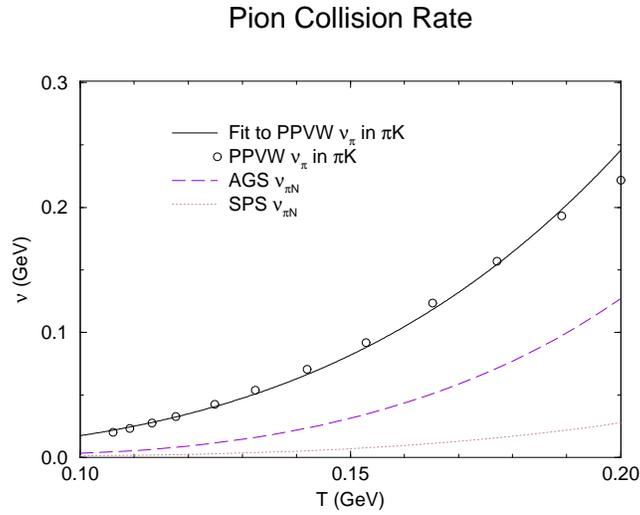}
\end{center}
\caption{\label{fig-app-collision-1} Pion collision rates
  $\nu=\tau^{-1}_{coll}$ [GeV] in a pion-kaon-nucleon gas
versus temperature T [GeV].}
\end{figure}

Using $\mu_b(T)$ from the previous section we can
evaluate the above integral numerically. The results are shown in
Fig.\ref{fig-app-collision-1}, for the $\pi\pi,\pi K$ rates combined (dots and fitted
curve) and for its $\pi N$ component at AGS and SPS.
Total pion collision time is then given by

\begin{eqnarray}
\label{eq-app-collision-0}
\tau_{collision}(T) = (\Gamma^{\pi}_{\pi\pi}(T) +
\Gamma^{\pi}_{\pi N}(T))^{-1}
\end{eqnarray}

\begin{figure}[t]
\begin{center}
\includegraphics[width=3.8in]{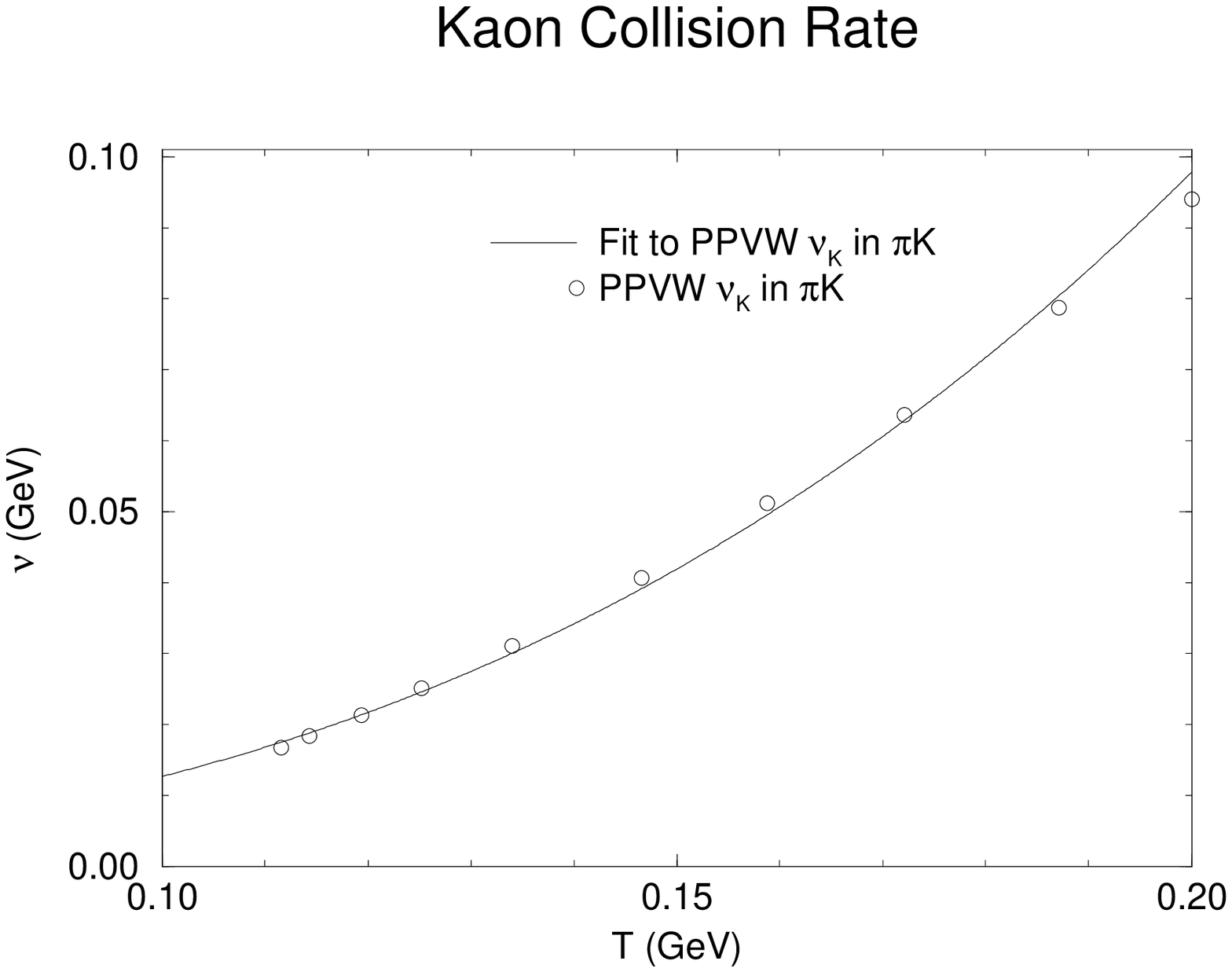}
\end{center}
\caption{\label{fig-app-collision-2} Kaon collision rates}
\end{figure}

For kaons we simply take the $\pi K$ rate from \cite{PPVW_rates} which we show
in Fig.\ref{fig-app-collision-2}. We have ignored smaller $KN$ 
collision rates: therefore  (as also noted in\cite{PPVW_rates}) 
 we do not   distinguish the rates for kaons and anti-kaons.

\begin{figure}[t]
\begin{center}
\includegraphics[width=3.8in]{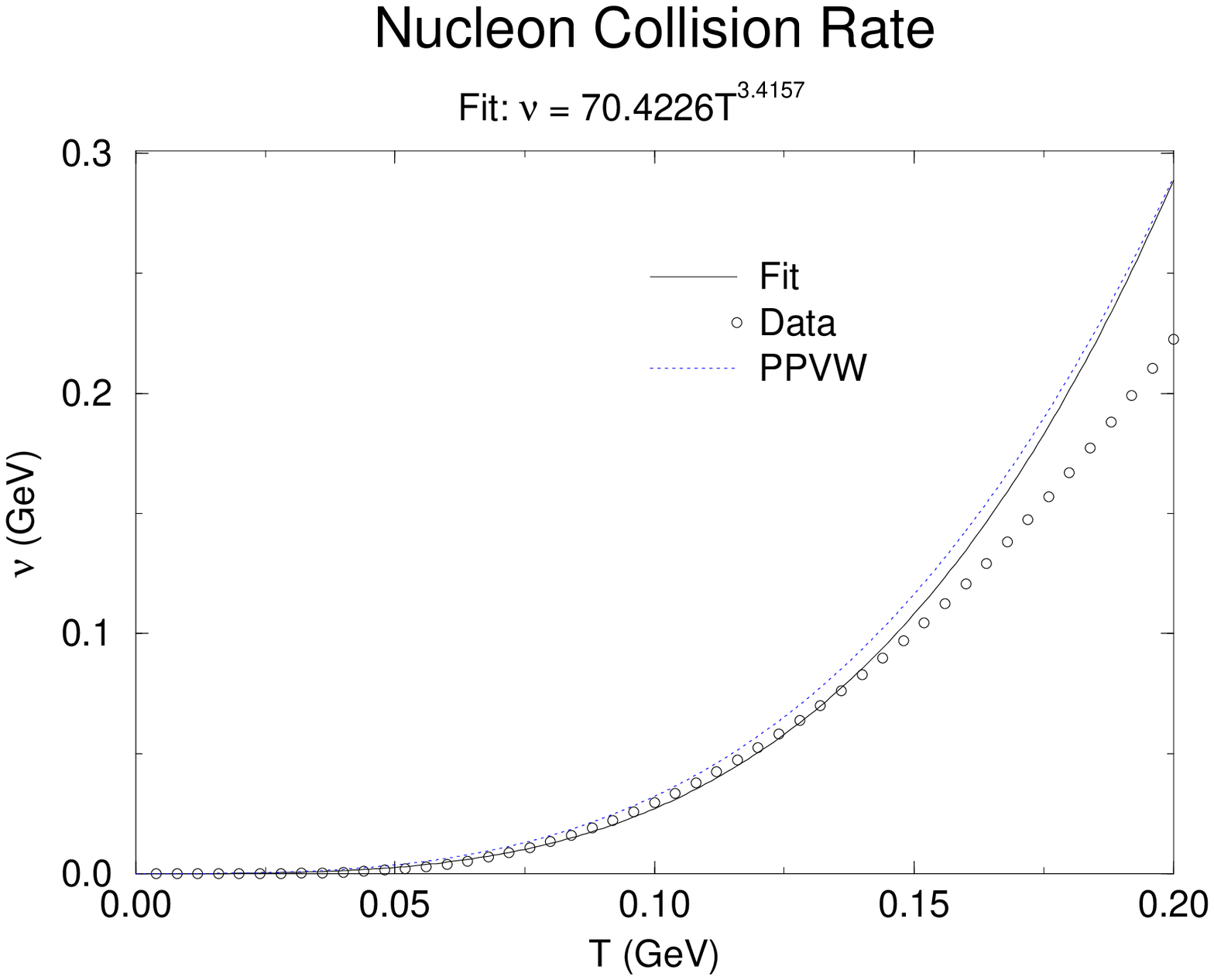}
\end{center}
\caption{\label{fig-app-collision-3} Nucleon
  collision rates. ``Data'' refers to our numerical results, and
  ``Fit'' refers to a fit to our results using the data points below
  0.15 GeV}
\end{figure}

For nucleons, $\pi N$ interaction is the dominant process\cite{PPVW_rates}
and we have the expression:

\begin{eqnarray}
\label{eq-app-collision-9}
\Gamma^{N}_{\pi N}(T) =
{1\over{\int{d^3{\bf p}_N \over (2\pi)^3} {1\over e^{(E_N-\mu_b)/T} + 1}}}
\int{d^3{\bf p}_N \over (2\pi)^3}\int{d^3{\bf p}_\pi \over (2\pi)^3}
{2\over e^{E_\pi/T} -1}{1\over e^{(E_N - \mu_b)/T} +1}
\cr
\sigma_{\pi^+p}\left[(p_\pi + p_N)^2\right]
\left| {{\bf p}_\pi\over E_\pi} - {{\bf p}_N\over E_N} \right|
\end{eqnarray}

noting that $g_\pi = 3$.  It turns out that due to the almost
factorizable nucleon density inside the main integral which almost cancels the
denominator in front, the effect of $\mu_b$ is almost negligible.
In Fig.\ref{fig-app-collision-3} we present the common rate for the AGS and SPS and compare with
the SPS nucleon rate from \cite{PPVW_rates}.

  The remaining issue is what value of the ratio in
  the condition (\ref{eq_cond}) one should  use in order to optimize
  the
surface. Consider for example the simplest case in which 
expansion/reactions proceed with the same rates for some time, the
densities are $n\sim exp(-\tau/\tau_{exp})$. Plugging it into the rate  
and integrating from freeze-out to infinity one gets the number of
collisions left over to be $\sim (1-exp(-\xi))$. As we want to cut
roughly in the middle of the last collision, one may think the optimal
point is close to $\xi=1/2$. Our checks with cascades (see below)
confirm this choice, although one may in future improve on this point.

\begin{figure}[t]
\begin{center}
\vskip 0.5in
\epsfxsize=3.8in
\centerline{\epsffile{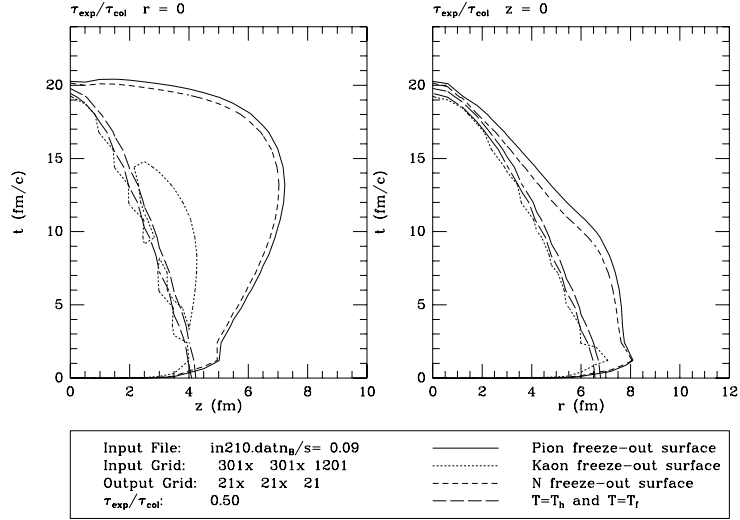}}
\end{center}
\caption{\label{fig-b-spect-1} Freeze-out surfaces for 11.6A GeV Au+Au}
\end{figure}

\begin{figure}[t]
\begin{center}
\vskip 0.5in
\epsfxsize=3.8in
\centerline{\epsffile{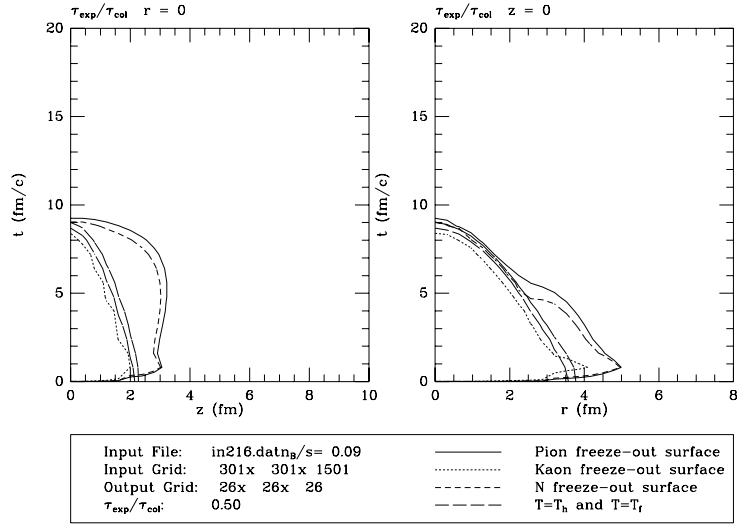}}
\end{center}
\caption{\label{fig-b-spect-5} Freeze-out surfaces for 14.6A GeV Si+Al}
\end{figure}

\begin{figure}[t]
\begin{center}
\vskip 0.5in
\epsfxsize=3.8in
\centerline{\epsffile{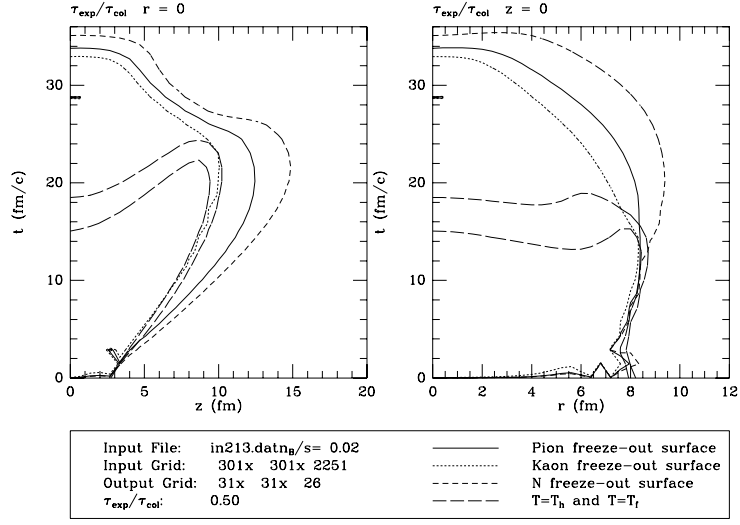}}
\end{center}
\caption{\label{fig-b-spect-6} Freeze-out surfaces for 160A GeV Pb+Pb}
\end{figure}

\begin{figure}[t]
\begin{center}
\vskip 0.5in
\epsfxsize=3.8in
\centerline{\epsffile{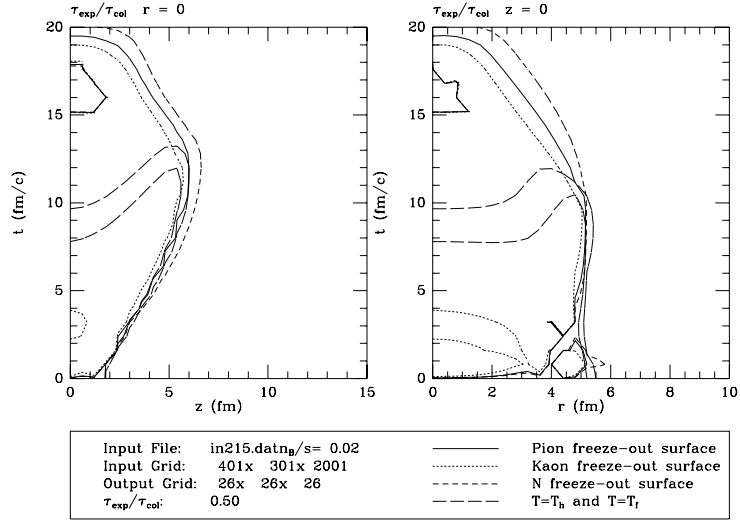}}
\end{center}
\caption{\label{fig-b-spect-7} Freeze-out surfaces for 200A GeV S+S}
\end{figure}

Using these rates, and the condition (\ref{eq_cond}), we  determine
the freeze-out (3-d) surface. Several representative cases are shown
in Figs.\ref{fig-b-spect-1} to \ref{fig-b-spect-7}, shown as a section
by time t - longitudinal coordinate z plane (at transverse
coordinate r=0) and the t-r plane (z=0).
We have already commented about dependence on the particle kind above.
Note also a significant difference between heavy and light ions.
As expected,  
one finds that the larger is the system, the lower is $T_{th}$ on this surface.

Furthermore, the shape of the freeze-out surface is very different
from simple isotherms. It means that
there is a significant variation of this temperature
over the surface itself:
in order to find the coolest pion gas, one should
look at the very center of
central collisions of heaviest nuclei at highest available energy!

  Finally, after elements of the  the freeze-out 
surface are determined (from hydro
solution plus kinetic condition discussed above)  by (3-d)
  triangularization (see appendix),
 the HKM program generates secondaries using the
Cooper-Frye formula\footnote{Although this formula is conserving
  energy and is widely used, there is still a well known problem with
  it when applied to the $space-like$ part of the freeze-out surface:
it includes also particles which move toward the excited system, which
 would be re-absorbed. Possible improvements are discussed
e.g. in recent work \cite{Czernai}: we have not included those in HKM.
}

\begin{eqnarray}
\label{eq-b-spect-1}
E{dN\over d^3{\bf p}} =
{1\over (2\pi)^3}\int_{\sigma_f}{p^\mu d\sigma_\mu\over e^{p\cdot u/T}
\pm 1}
\end{eqnarray}

where the integral is taken over
freeze-out surface, with  $T_{th},\mu_{th},u_\mu$ changing from
point to point.

 The last step of the
HKM  is
the decay  of all resonances (and stable particles, if needed) into
the final secondaries. The standard
output, like from other event generators, includes information about
particle momenta, the time and  place of their production, and the
parent resonance (if they come from a decay).

\section{ Further hydro results:  the radial flow}

  In the previous section we have shown that the improved thermal
  freeze-out condition lead to huge 
difference  compared with simple  isotherms.
  For the same fixed value of the
$\tau_{exp}/\tau_{coll}  $
one gets very different conditions for different A, also for
 large and small y (the central
region cools further).
 These observations provide natural resolutions to the puzzling
observations related to strong A- and y-dependence of the flow
discussed above.

The key point here is as follows:
although these modifications does not significantly prolong total
lifetime, it significantly
increases  the lifetime of hadronic phase. It is important for flow,
because it is    the part of the evolution path at which the matter is 
most ``stiff'' (have larger $p/\epsilon$). 
 Thus improved freeze-out leads to a significant ``extra push'',
and explains strong flow.

\begin{figure}[t]
\vskip 0.5in
\epsfxsize=3.8in
\centerline{\epsffile{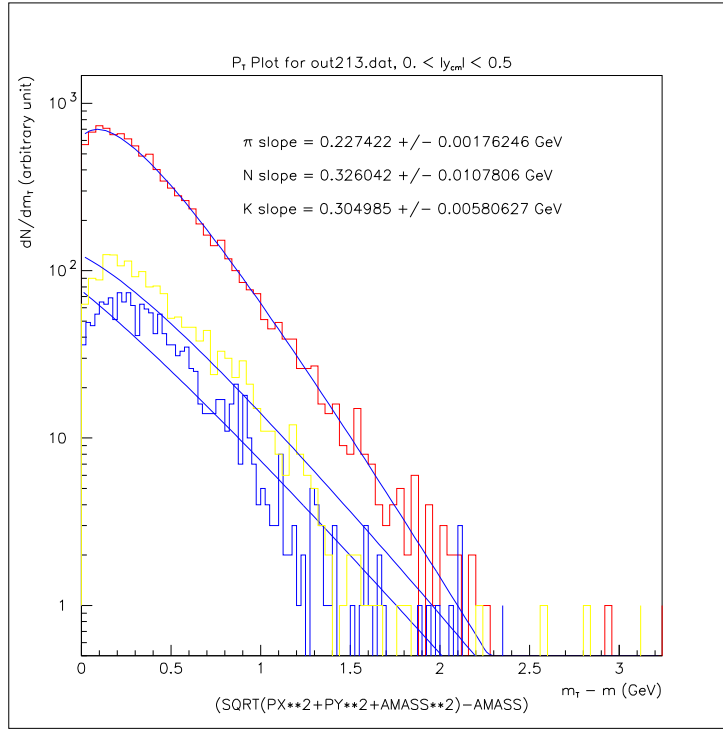}}
\caption{\label{ptspectra} Typical  hydro output and 
fit of the $m_t$ distributions for pion, kaon and protons, for central
PbPb collisions at 158 GeV A, at central rapidity $|y|<.5$ .
 }
\end{figure}

The typical $p_t$ spectra for $\pi,K,N$ we obtain (after resonance
decays)
are shown  in Fig.\ref{ptspectra} (a), together with their fixed-slope fits.

\begin{figure}[t]
\begin{center}
\vskip 0.5in
\epsfxsize=3.8in
\centerline{\epsffile{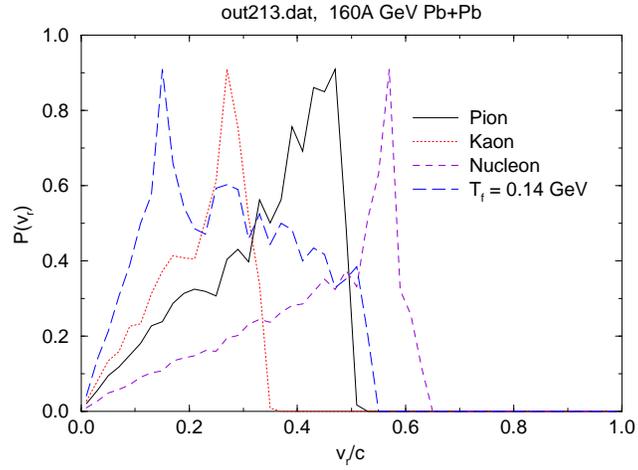}}
\end{center}
\caption{\label{out213vr} Transverse velocity distribution over the
  various freeze-out surfaces for 160A GeV Pb+Pb collision  }
\end{figure}

\begin{figure}[t]
\begin{center}
\vskip 0.5in
\epsfxsize=3.8in
\centerline{\epsffile{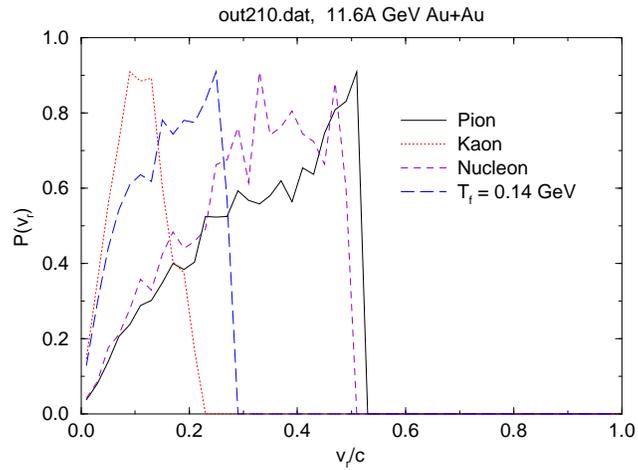}}
\end{center}
\caption{\label{out210vr} Transverse velocity distribution over the
  various freeze-out surfaces for 11.6A GeV Au+Au collision  }
\end{figure}

 In Figs.\ref{out213vr}
and \ref{out210vr} we
show the distribution over transverse velocities calculated over all
matter element on the freeze-out surfaces. We show only heavy ions,
for AGS and SPS energies. 
The distributions always
have sharp peak at their right end, which is more pronounced at SPS.
Its position depend significantly on the particle type, reaching
as high speak as $v_t=0.6$
for N at SPS. Note dramatic difference with the isotherms $T=.14 GeV$
which were used in many previous works: for them there is also a peak,
but for much smaller  $v_t\approx .17$, plus a shoulder toward larger values.
This difference is much smaller for medium ions (not shown).

\begin{figure}[t]
\begin{center}
\vskip 0.5in
\epsfxsize=3.8in
\centerline{\epsffile{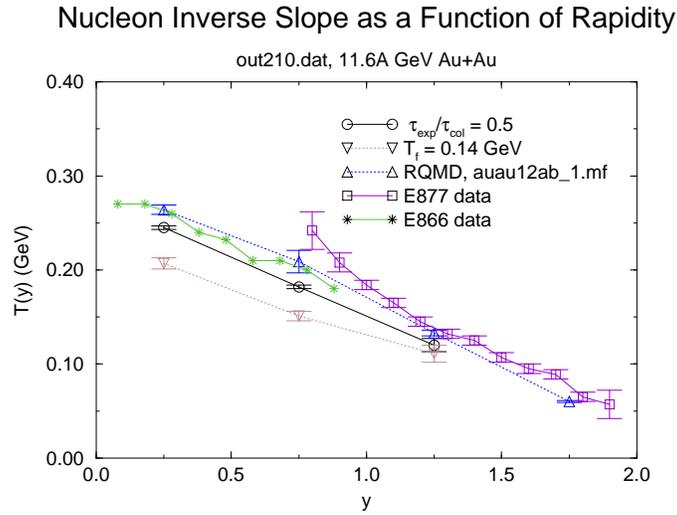}}
\end{center}
\caption{\label{flowslopes1} Nucleon slope parameters for 11.6A GeV Au+Au  }
\end{figure}

\begin{figure}[t]
\begin{center}
\vskip 0.5in
\epsfxsize=3.8in
\centerline{\epsffile{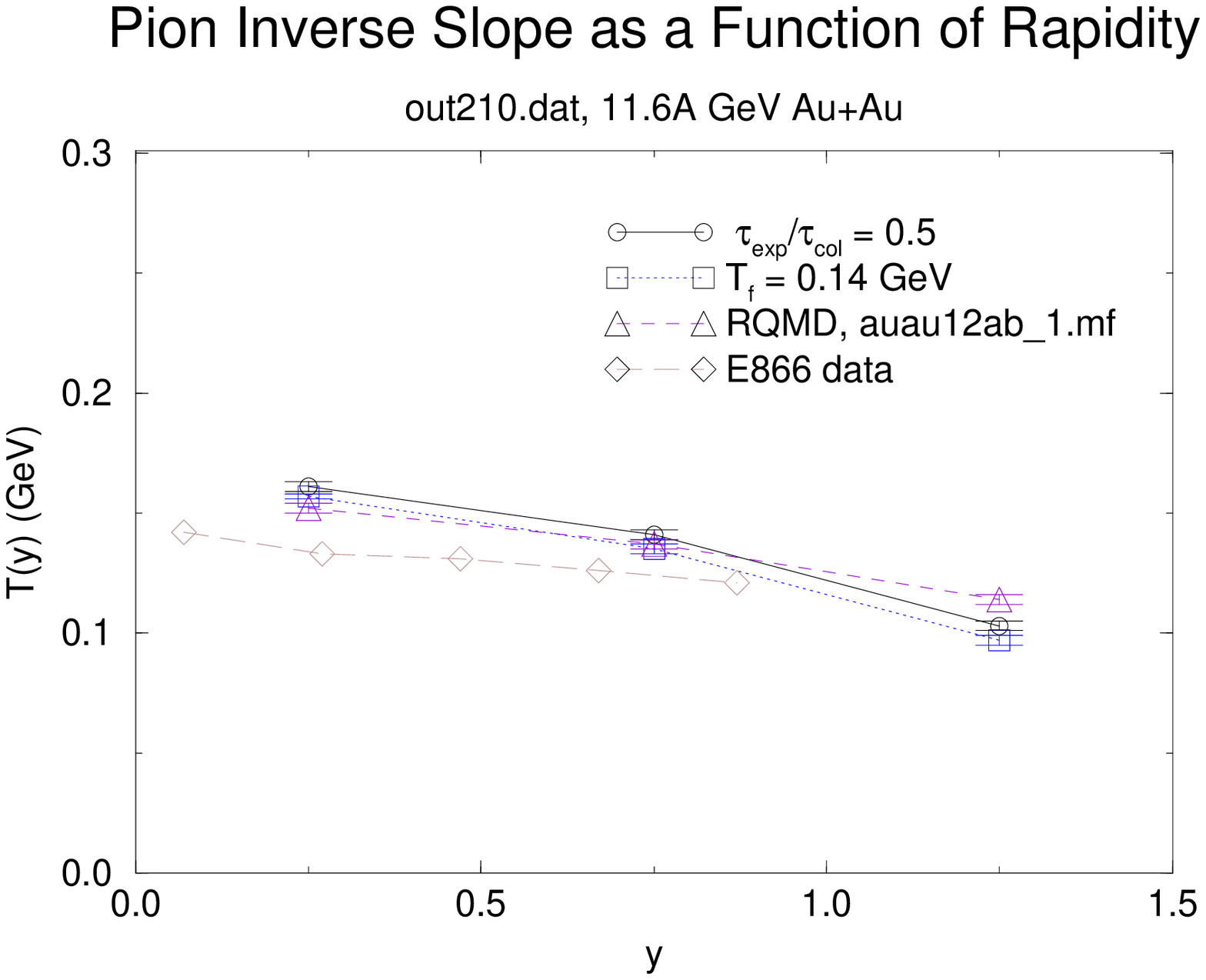}}
\end{center}
\caption{\label{flowslopes1a} Pion slope parameters for 11.6A GeV Au+Au  }
\end{figure}

\begin{figure}[t]
\begin{center}
\vskip 0.5in
\epsfxsize=3.8in
\centerline{\epsffile{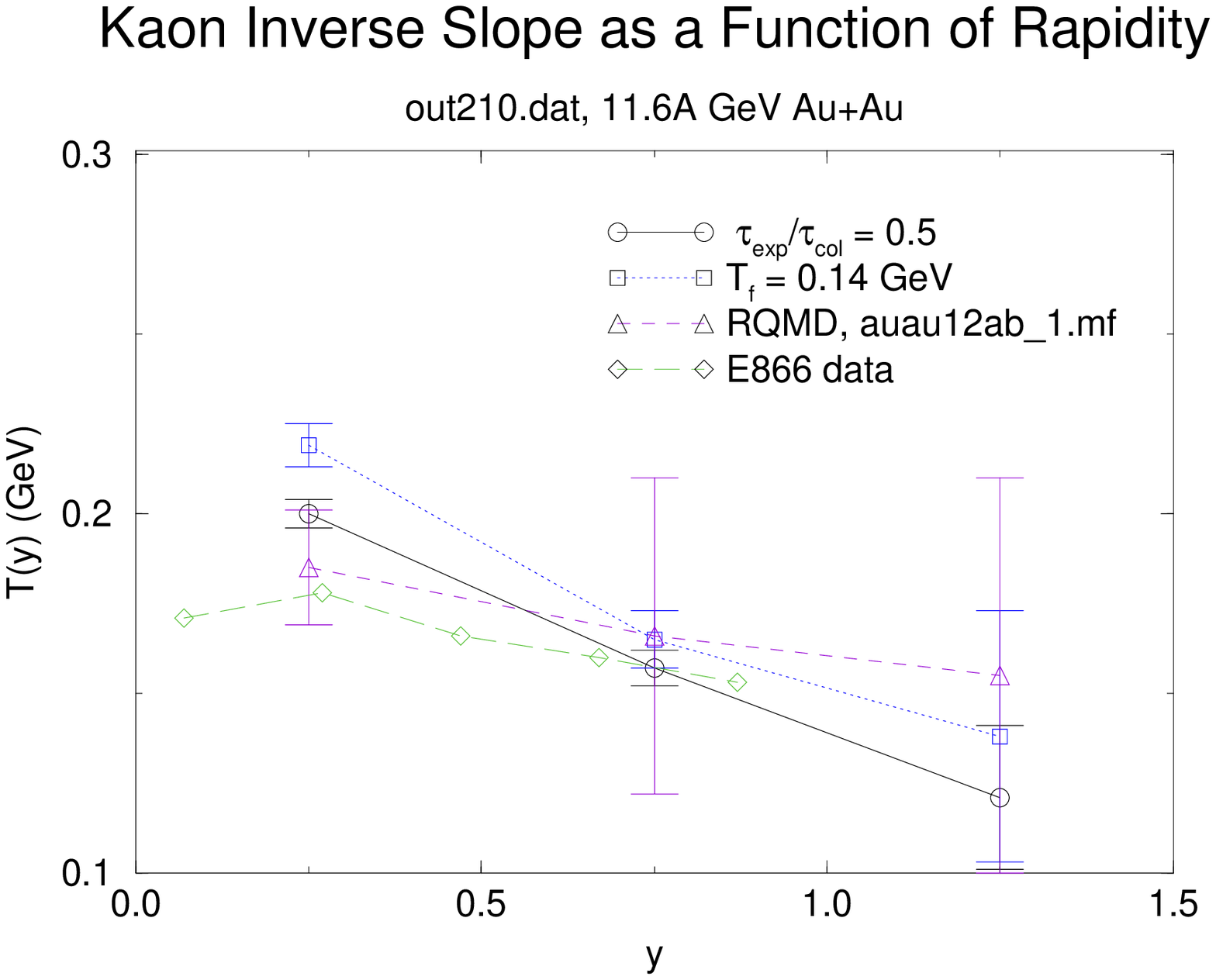}}
\end{center}
\caption{\label{flowslopes1b} Kaon slope parameters for 11.6A GeV Au+Au  }
\end{figure}

\begin{figure}[t]
\begin{center}
\vskip 0.5in
\epsfxsize=3.8in
\centerline{\epsffile{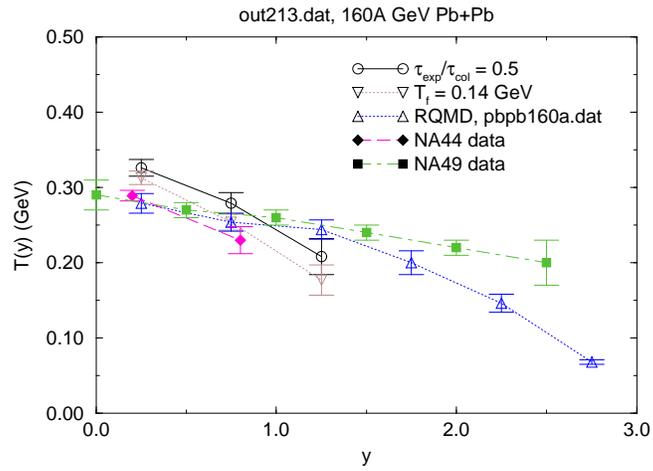}}
\end{center}
\caption{\label{flowslopes2} Nucleon slope parameters for 158A GeV Pb+Pb  }
\end{figure}

\begin{figure}[t]
\begin{center}
\vskip 0.5in
\epsfxsize=3.8in
\centerline{\epsffile{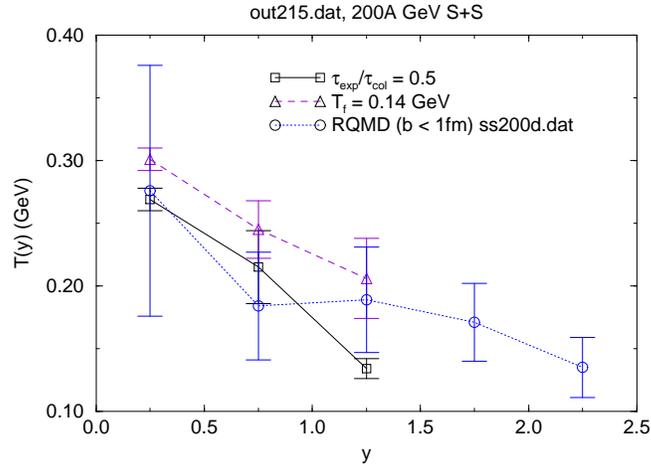}}
\end{center}
\caption{\label{flowslopes3} Nucleon slope parameters for 200A GeV S+S  }
\end{figure}

\begin{figure}[t]
\begin{center}
\vskip 0.5in
\epsfxsize=3.8in
\centerline{\epsffile{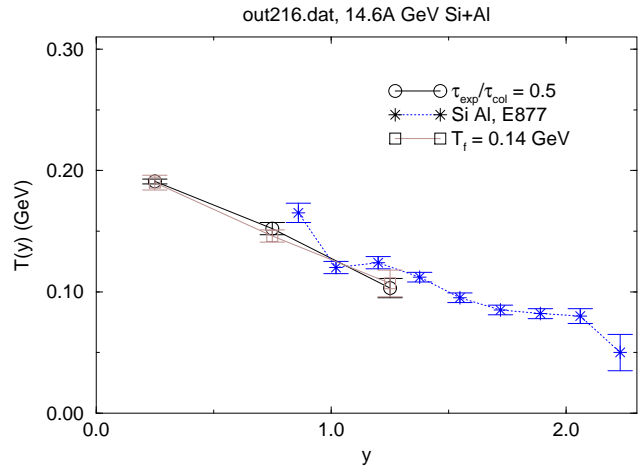}}
\end{center}
\caption{\label{flowslopes4} Nucleon slope parameters for 14.6A GeV Si+Al  }
\end{figure}

In Figs.\ref{flowslopes1} to \ref{flowslopes4}
 we show how this translates into
 the observable quantity, the   $m_t$
 slopes    $\tilde T(y)$. Recall that they include effect of
 freeze-out temperature, flow plus resonance decays, and we show them
as a function of rapidity y.
 We show 4 cases: AuAu at 11 GeV/N, PbPb at 158 GeV/N SiAl at 14.6
 GeV/N and S S at 200 GeV/N.  
    In all cases we compare our results with the experimental data
    available, as well as with the
 RQMD (which was
   obtained from standard output files and fitted
in the same way as the HKM ones).

  For AuAu data at AGS  Fig.\ref{flowslopes1} one can see, that RQMD reproduces slopes very 
accurately, while our results slightly under-predict the flow. However,
it is precisely how it should be, because this version of RQMD has
been tuned with a
repulsive baryon-induced potential, on the top of the pure cascade. We
have checked that the version without potential
 gives smaller flow, and agrees with
our results very well.  At the same time, the 
results following from ``naive'' freeze-out
with $T_f=140 MeV$ is way below.

  Fig.\ref{flowslopes2}, showing  PbPb at 158 GeV/N, look very
  similar to Fig.\ref{flowslopes1}. This feature however must be a mere coincidence,
since both the EOS\footnote{Due to completely different
 matter composition: the $\pi/N$ ratio
different by factor 5.} and  the space-time picture are quite
different. Apart from obviously rather different longitudinal motion,
at AGS the transverse velocity is gained gradually  in time (due to
about
constant $p/\epsilon$, or acceleration, while at SPS our hydro
solution
clearly display appearance of a ``burning wall'' regime, at which
 most of acceleration occurs. Note that nevertheless our results agree
 with
data and RQMD in this case as well, for the $same$ $\tau_{exp}/\tau_{coll}=.5$.
 This agreement is very non-trivial.

  For comparison,  let us now discuss lighter ions. An example is shown in 
 Fig.\ref{flowslopes3}, for S S at 200 GeV/N, and one can see from
 it that our results over-predict flow in the central region $y\approx
 0$. Although for light ions HKM predicts shorter
lifetime of the hadronic phase and smaller flow ($ \tilde T_N(y)$
about 30 percent lower), the data (and RQMD) show
that this drop should in fact be larger. It is hardly surprising
to see 
 that for medium ions the  HKM  (and probably hydro-based models in general) 
 are  less accurate. 

  Let us finally stress that we have not attempted any fine tuning of the
  parameters used. The main ingredients, EOS and freeze-out parameter 
$\xi=1/2$ were fixed rather early and not modified when hadron
spectra/slopes
were calculated. Clearly one can do it and get better agreement.
In this work our main objective was to test crudely the systematics of
the flow discussed in the introduction (and, of course, its magnitude).

  Finally, a comment on agreement  with RQMD is in order here. We emphasized
  above that its EOS is similar to ours for AGS domain, in which both
 represent the resonance gas: but how can both agree at SPS energies,
 where our EOS has the notorious softness due to the QCD phase transition?  
 In fact, RQMD has its own reasons for changing its EOS to larger 
 ``softness'':  at SPS
 conditions at early times the energy is stored no longer in
 resonances, but
  in (longitudinally
 stretched) strings. Naturally, those make little pressure in
 the transverse direction. 
   
\begin{figure}[t]
\begin{center}
\vskip 0.5in
\epsfxsize=3.8in
\centerline{\epsffile{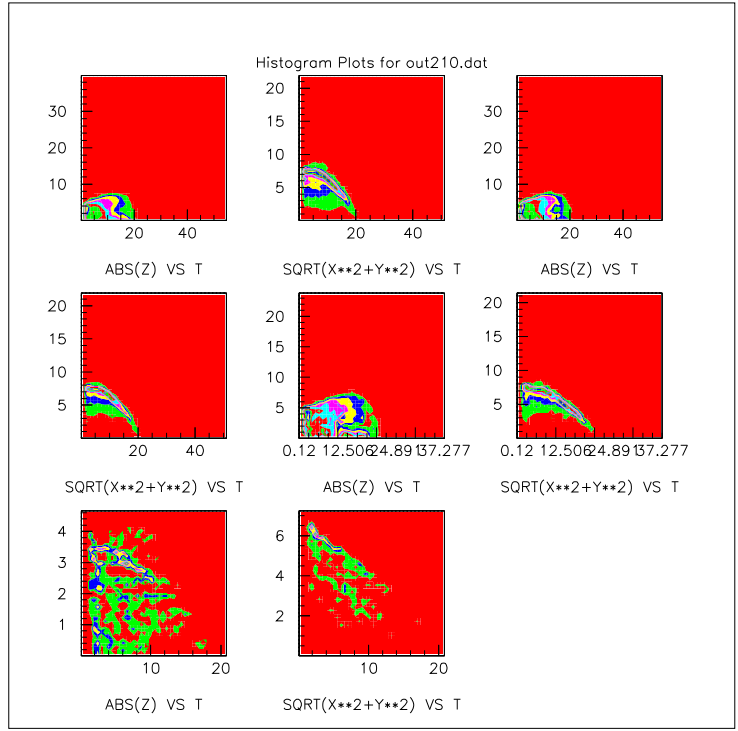}}
\end{center}
\caption{\label{fig-b-appl1-4} Space-time distributions for 11.6A GeV
  Au+Au, hydro. Eight pictures are projections of the emission points
on the z,t and r,t planes,
for all secondaries, pions, nucleons and kaons,
subsequently. Resonance decays are included.}
\end{figure}

\begin{figure}[t]
\vskip 0.5in
\epsfxsize=3.8in
\centerline{\epsffile{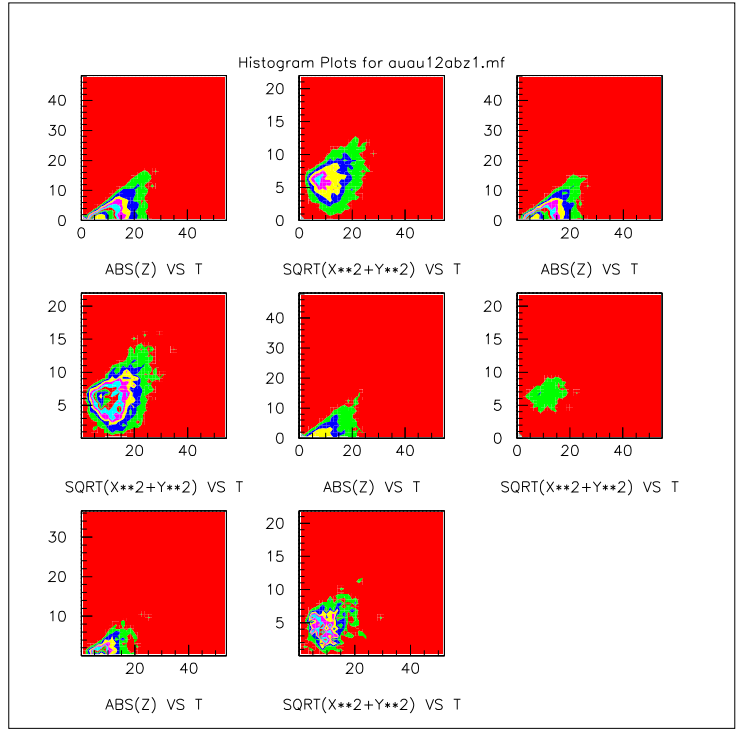}}
\caption{\label{fig-b-appl1-5} Same distributions as the previous
  figure (also 11.6A GeV
  Au+Au) but for RQMD output file.  }
\end{figure}

  By no means we want to create an
  impression that our model and RQMD are to a large extent identical. The
  magnitude and various dependences of the flow we discussed in this
  work are important observables, but even those give only partial  
   information on the space-time picture of the collision.
  Looking at these results more closely, one however finds significant
  differences here, which  
should affect the ``freeze-out sizes'' extracted by pion interferometry
 analysis. This statement is illustrated in Figs.\ref{fig-b-appl1-4}
 and \ref{fig-b-appl1-5},
 comparing distributions in the points of the last interaction in our
 model and RQMD. One can see from it that although
the average sizes generally agree (and thus flow velocities), their $dispersions$
(relevant for interferometry) are rather different.
With better data coming, one would be soon able to address this aspect
as well.

\section{Summary and discussion}

  In this work we have developed next-generation
 hydro-kinetic model for heavy ion
  collisions. Although most of the ideas in it are not new, we
  believe they are now brought together in an economic and practical
  way.

  Compared to previous hydro-based models in literature we have included
a number of improvements: (i) realistic EOS including the QCD phase
transition
together with the
effect of baryons; (ii) more realistic 
 ``local'' freeze-out
  condition, which is based directly on kinetics of re-scattering;
(iii) decay of all resonances in the final state, etc.

  Our main focus was on new data on radial flow, including its
magnitude, y, A
  and s-dependence. We have found that our model in general 
  reproduces it well enough. This shows that the lattice-based EOS
(which is very soft in the transition region, as we repeatedly emphasized)  
is in fact consistent with flow data. This is our main result.

  The crucial observation which was important for this success is point
  (ii). It
leads to a very simple property of the freeze-out: 
the larger is the size of the system, the cooler   the  matter
at the end becomes. (We have found it  surprising that
this effect 
 was overlooked before.) Clearly, deeper cooling for larger A should be
 seen in many different ways, and we look forward other ways of
 testing it.

  One may further ask, whether data can 
restrict the EOS. We have not attempt to quantify this
in the present work, and only note that for EOS  
  $without$ the QCD phase transition 
(e.g.  a resonance gas,
 with  $p/\epsilon\approx constant$ discussed above) the magnitude of the
flow for SPS is indeed too large, and the expansion time too short.
Clearly further studies are needed to
clarify these issues. Natural extension is
 discussing 3+1 hydro at non-zero impact parameters, leading to
dipole/elliptic components of the flow.
  
   Another obvious way to proceed is to
 calculate the HBT radii and compared it with data. We have already
mention that $dispersions$ of the emission time/positions in our model
are quite different from those resulting from the RQMD. 

   Clearly only a small fraction of  data is considered in this work. To
  facilitate further use of the model,
  we plan to deviate from the usual scenario in which only the basics
  formulae plus $some$ results are presented in the paper, and 
plan to provide the source code/output files,
 in the same form as event generators
do. We hope it will prompt the experimentalist to use it widely, revealing
 in wider scope its agreement and disagreement with particular data.

\appendix
\section{Triangulation of the Freeze-Out Surface}
\label{chap-app-triangle}
The surface element on a 3-surface of space-time is given by\cite{mtw}

\begin{eqnarray}
\label{eq-app-triangle-1}
d^3s_\mu = \epsilon_{\mu\alpha\beta\gamma}{\partial x^\alpha\over\partial a}
{\partial x^\beta\over\partial b}{\partial x^\gamma\over\partial c}
da db dc
\end{eqnarray}

where $a, b, c$ are the coordinates on the 3-surface and
$\epsilon_{\mu\alpha\beta\gamma}$ is the totally
antisymmetric Levi-Civita tensor.
For our purposes, due to the cylindrical symmetry of our model, we
need to express a finite 3-surface element of a freeze-out surface in
terms of its corner points.  More specifically we want to find
$d^3s_\mu$ for a triangle defined in $(z,r,t)$ space with corner
points $(z_i,r_i,t_i)\ \ \ i=1,2,3$. It can be shown that up to a sign,

\begin{eqnarray}
\label{eq-app-triangle-2}
d^3s_\mu = {1\over 2}rd\theta
\left(
\left|\matrix{z_1 & r_1 & 1 \cr z_2 & r_2 & 1 \cr
  z_3 & r_3 & 1 \cr}\right|,
\left|\matrix{t_1 & z_1 & 1 \cr t_2 & z_2 & 1 \cr
  t_3 & z_3 & 1 \cr}\right|\cos\theta,
\left|\matrix{t_1 & z_1 & 1 \cr t_2 & z_2 & 1 \cr
  t_3 & z_3 & 1 \cr}\right|\sin\theta,
\left|\matrix{r_1 & t_1 & 1 \cr r_2 & t_2 & 1 \cr
  r_3 & t_3 & 1 \cr}\right|
\right)
\cr
\end{eqnarray}

The sign would have to be determined by choosing a direction for the
normal of the surface which points outward from the hotter interior of
the surface.  The output file of the hydro program gives
$(t,z,r,\epsilon,\xi=\tau_{exp}/\tau_{col},...)$ at each point of the output grid (which is
typically of size $25 \times 25 \times 25$ in $(z,r,t)$).  To triangulate the freeze-out
surface, we pick a cell and check to see whether
$\xi$ (or $\epsilon$ if we want a freeze-out surface of constant
temperature) on its vertices are above or below the freeze-out value
$\xi_f$ (or $\epsilon_f$).  By interpolation we can determine the
intersections (if any) between the freeze-out surface and the edges of
the cell.  Once the intersections are found, we find the center point
of these intersection points and connect it to two adjacent
intersection points to form a triangle. Continuing this process for
all the cells we obtain the desired triangulation of the freeze-out
surface.

\par

\clearpage

\begin{figure}[t]
\vskip 0.5in
\epsfxsize=3.5in
\centerline{\epsffile{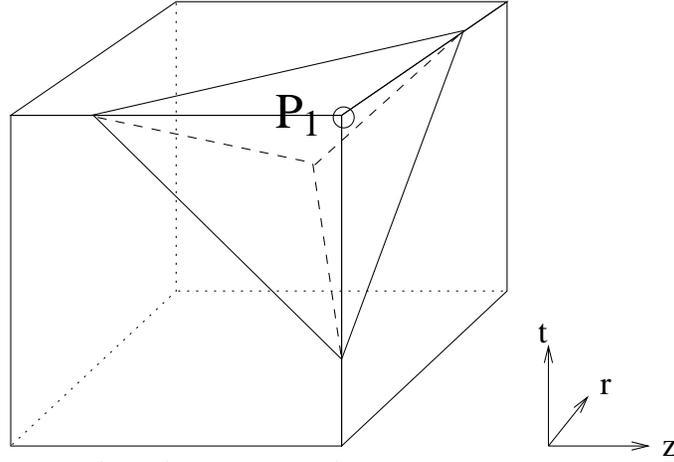}}
\caption[Triangulation of the freeze-out
  surface within a single cell]{\label{fig-app-triangle-1} Triangulation of the freeze-out
  surface within a single cell.  Here $P_1$ is the only vertex with
  $\xi > \xi_f$ while all other vertices have $\xi < \xi_f$.  This
  cell will yield 3 triangles upon triangulation.}
\end{figure}

\end{document}